\documentclass[11pt, a4paper]{article}
\usepackage{amsfonts}
\usepackage{amsmath}
\usepackage{amssymb}
\usepackage{amsthm}
\usepackage{rotating}
\usepackage{color}
\usepackage[table]{xcolor}
\usepackage{endnotes}
\usepackage{booktabs}
\usepackage{threeparttable}
\usepackage{longtable}
\usepackage{hyperref}
\usepackage{float}
\usepackage{natbib}
\usepackage{bm}
\usepackage{mathtools}
\usepackage{amssymb, longtable}
\usepackage{pdflscape}
\usepackage{arydshln}

\definecolor{LightGray}{gray}{0.95}
\definecolor{LightBlue}{rgb}{0.85,0.93,1.00}
\providecommand{\U}[1]{\protect\rule{.1in}{.1in}} \textwidth=420pt
\RequirePackage{amsopn} \RequirePackage{amsfonts}
\RequirePackage{amsthm}

\DeclareMathOperator*{\argmin}{\arg\!\min}
\newcommand{\R}{\mathbb{R}}
\begin{document}

\title{Estimation of income inequality from grouped data}
\author{Vanesa Jorda\thanks{%
Corresponding author: jordav@unican.es. Department of Economics, University of Cantabria. Avda. de los Castros, s/n, 39005, Santander (Spain)}, Jose Maria Sarabia\\
    \textit{Department of Economics, University of Cantabria}\\ \and 
  Markus J\"antti\\
  \textit{Swedish Institute for Social Research, Stockholm University}}
\maketitle

\begin{abstract}

Grouped data in form of income shares have been conventionally used to estimate income inequality 
due to the lack of availability of individual records. Most prior research on economic inequality relies on lower bounds of inequality measures in order to avoid the need to impose a parametric functional form to describe the 
income distribution. These estimates neglect income differences within shares, introducing, therefore, 
a potential source of measurement error. The aim of this paper is to explore a nuanced alternative to estimate 
income inequality, which leads to a reliable representation of the income distribution within shares. We 
examine the performance of the generalized beta distribution of the second kind and related models to 
estimate different inequality measures and compare the accuracy of these estimates with the nonparametric
lower bound in more than 5000 datasets covering 182 countries over the period 1867-2015. We deploy two 
different econometric strategies to estimate the parametric distributions, non-linear least squares and generalised 
method of moments, both implemented in R and conveniently available in the package \texttt{GB2group}. 
Despite its popularity, the nonparametric approach is outperformed even the simplest two-parameter models. 
Our results confirm the excellent performance of the GB2 distribution to represent income data for a heterogeneous sample of countries, which provides highly reliable estimates of several inequality measures. 
This strong result and the access to an easy tool to implement the estimation of this family of distributions, we believe, will incentivize its use, thus contributing to the development of reliable estimates of inequality trends.

\end{abstract}


\\
\newpage
\section{Introduction}

The analysis of the income distribution has a venerable history within social sciences. Its evolution has been considered essential to explain not only the causes but also the potential consequences of inequality and poverty. The role of changes in the income distribution on different socio-economic aspects, such as growth, consumption or human capital formation, is widely documented in the literature (see e.g Barro, \citeyear{barro2000}; Krueger et al. \citeyear{krueger2006}). Much empirical research has also been directed at examining the temporal evolution and the geographical differences in poverty and inequality, considering as potential determinants of these two phenomena the family structure, medical progress or technological change, to mention a few (Deaton, \citeyear{deaton2013}; McLanahan and Percheski, \citeyear{mclanahan2008}). 

When individual records on personal or household income data were available, the estimation of the income distribution would be relatively simple, using the empirical distribution function. However, much of the existing scholarship on economic inequality has been plagued by a lack of individual data. This potential limitation is particularly severe for studies with large geographic coverage, which involve several countries at different points in time. Nevertheless, the periodic release of certain summary statistics of the income distribution has become increasingly common. The World Income Inequality Database (WIID), World Bank's PovcalNet or the World Wealth and Income Database (WID) are the largest cross-country databases that store grouped income/consumption data, typically including information on few income and population shares. This type of grouped data depicts sparse points of the Lorenz curve and hence, to estimate inequality measures, it is essential to define a method to link such points. 

Much of the academic literature on the estimation of income inequality from grouped data deploy nonparametric techniques to approximate the shape of the Lorenz curve. Linear interpolation of the income shares is the most common approach to construct the so-called empirical Lorenz curve from which relative inequality measures are obtained. With very few exceptions, the extant scholarship in the global distribution of income presents inequality trends based on this method (Bourguignon and Morrison \citeyear{bourguignon2002}; Milanovic, \citeyear{milanovic2011}; Lakner and Milanovic, \citeyear{lakner2013}; Ni\~no-Zarazua et al., \citeyear{nino2014}; Dowrick and Akmal, \citeyear{dowrick2005}). The broad popularity of this methodology is not only due to its simplicity, but also because it is argued that there is no need to impose any particular model to fit the empirical data. However, despite not explicitly, this approach rests on a predefined distributional model. Indeed, it assumes that all individuals within a particular quantile have the same level of income, which obviously does not represent the income distribution accurately. As a result, the actual level of inequality is underestimated and, consequently, relative inequality measures estimated within this framework are lower bound estimates (see, e.g., Kakwani,  \citeyear{kakwani1980}). 
 
Therefore, to obtain reliable estimates of inequality measures, we need to rely on a model which allows us to define more plausible assumptions on the income distribution within income shares. Due to its flexibility, some authors opted for kernel estimation because it avoids imposing a particular functional form for the income distribution (Sala-i-Martin, \citeyear{sala2006}). However, the performance of this approach seems to be extremely sensitive to the selection of the bandwidth parameter, which might lead to significant biases on the estimates of poverty and inequality measures (Minoiu and Reddy, \citeyear{minoiu2014}). Although parametric models seem to be a suitable alternative to nonparametric techniques for estimating income distributions (Dhongde and Minoiu, \citeyear{dhongde2013}), this approach has been barely used to estimate income inequality. The reason seems to be the need to make ex-ante assumptions on the shape of the distribution. If our choice is not a valid candidate to represent the income distribution, our estimates on inequality measures might be severely be affected by misspecification bias.

Despite this potential limitation, prior research suggests that the parametric approach outperforms other nonparametric techniques to estimate poverty indicators from grouped data (Dhongde and Minoiu, \citeyear{dhongde2013}; Bresson, \citeyear{bresson2009}). However, systematic empirical research on the effectiveness of parametric models to estimate inequality measures is surprisingly scarce. Although prior research points towards an excellent performance of parametric models to estimate inequality measures (Cowell and Metha, \citeyear{cowell1982}; Minoiu and Reddy, \citeyear{minoiu2014}), these evaluations rely on single case studies and a limited range of distributions, meaning that the findings should be treated with great caution. Robust empirical evidence on the reliability of parametric estimates would, therefore, cast valuable light on the relative merits of this approach to estimate income inequality from grouped data.


In this paper, we explore the implications of using parametric models to estimate income inequality from grouped data for 5570 datasets, which cover more than 180 countries over the period 1867-2015. Among the whole range parametric distributions, we direct our attention to the generalised beta distribution of the second kind (GB2) and its particular and limiting cases. Several distributions that belong to this family have been commonly used to estimate the income distribution from grouped data (Chotikapanich et al., \citeyear{chotikapanich2007}; Jorda et al., \citeyear{jorda2014}; Pinkovskiy and Sala-i-Martin, \citeyear{pinkovskiy2014}) because it is acknowledged to provide an excellent fit to income data across different periods and countries (Feng et al., \citeyear{feng2006}, Hajargasht et al., \citeyear{hajargasht2012}; Bandourian et al., \citeyear{bandourian2002}). We have compared the accuracy of the parametric estimates of different inequality measures to their corresponding lower bounds to examine whether the preference of this method over the parametric approach, traditionally observed in the literature, is justified on the empirical ground. Our results show that the nonparametric approach performs very poorly on the estimation of income inequality. The GB2 distribution is confirmed as the best candidate to estimate income distributions, although the special cases of this family also lead to accurate estimates, being, in virtually all cases, more reliable than the lower bound estimates. Even for bimodal income distributions, which are clearly misrepresented by the GB2 distribution, we do not find evidence that supports the preference of the lower bound approximation of inequality measures over parametric models. 

This analysis confirms, therefore, that a common failure in much of the research on global inequality is the tendency to avoid using parametric functional forms. Of those studies that consider parametric models, most rely on simple two- or three-parameter distributions. Jorda and  Ni\~no-Zarzua (\citeyear{jorda2016}) is the only study that considered the GB2 distribution to estimate the global distribution of income. Country-specific applications are more common, but still scarce (see Burkhauser et al., \citeyear{Burkhauser2012}; Jenkins et al., \citeyear{Jenkins2011}; Feng et al. \citeyear{feng2006}). The lack of interest in this distribution, we believe, might be largely attributed to the fact that the efficient estimation of this model is far from straightforward. With the aim to make the GB2 distribution approachable and to incentivize its use, our estimation procedure implemented in R (R Development Core Team, \citeyear{team2013r}) is conveniently available in the R package \texttt{GB2group}.

In the following section, we introduce the notation and describe how the grouped data have been generated before presenting a summary of the GB2 distribution and its related models. The subsequent section discusses the estimation strategy based on nonlinear least squares (NLS) and the generalised method of moments (GMM) in a context of limited information. Thereafter, we compare the survey Gini index with both the so-called lower bound of inequality and the estimates based on parametric functional forms. We also present some results for model competition among different functional forms of the GB2 family to assess their performance to estimate the Gini coefficient. We make use of individual records to examine the robustness of the results to inequality measures that are more sensitive to the lower part of the distribution. We use Monte Carlo simulation to compare the performance of the parametric approach and the lower bound estimates to evaluate inequality levels of bimodal income distributions. We conclude the paper by considering the practical implications of our study.

\section{Estimation of income inequality from grouped data}

The use of grouped data to estimate income inequality has become increasingly a popular because individual records from surveys are not usually accessible for long periods of time, especially in developing countries. The public availability of large cross-country datasets, such as the WIID (UNU-WIDER, \citeyear{wiid3.4}) and PovcalNet, has motivated the use of this kind of data for the analysis of distributional issues. These databases gather information on summary statistics of the income distribution, typically the mean and few income shares.

Before going any further, it is crucial to understand how the grouped data has been generated. Let \textbf{x} be an $i.i.d.$ random sample of size $N$ from a continuous income distribution $f(x;\boldsymbol\theta)$ defined over the support $H = [0, \infty)$, where $\boldsymbol\theta \in \boldsymbol\Theta \subseteq \R ^k,$ being $\boldsymbol\Theta$ the parameter space. Assume that $H$ is divided into $J$ mutually exclusive intervals $H_j = (h_{j-1}, h_j] , j= 1, \dots, J$. We denote by $c_j= \sum_{i=1}^N\bm 1_{(h_{j-1}, h_j]}(x_i)x_i/\sum_{i=1}^Nx_i, j=1,\dots, J$ the proportion of total income held by individuals in the $j^{th}$ interval and the cumulative proportion by $s_j= \sum_{k=1}^j c_k$. Let $p_j = \sum_{i=1}^N\bm 1_{(h_{j-1}, h_j]}(x_i)/N , j=1,\dots, J$ denote the frequency of the sample \textbf{x} in the $j^{th}$ interval and $u_j= \sum_{k=1}^j p_k$ the cumulative frequency. According to this scheme, income shares ($s_j, j=1, \dots, J$) are ordinates of the Lorenz curve corresponding to the abscissae $u_j, j=1, \dots, J$.

The Lorenz curve informs about the proportion of income accruing to each cumulative share of the population, once incomes are arranged in increasing order. This curve is scale independent, hence changes the unit of measurement of the income variable, for instance, from dollars to thousand dollars, have no impact on the shape of the curve. Minimum inequality is observed when $s_j = u_j, j=1, \dots, J$, so that the Lorenz curve corresponds to the diagonal from the origin to the point (1, 1), which is known as the egalitarian line. The Lorenz curve is a powerful tool to compare and order distributions according to their inequality levels. If the Lorenz curve of one distribution lies nowhere below and somewhere above the curve of another distribution, the former distribution would be declared as less unequal than the later. 

To construct the Lorenz curve with the available information on income shares, we should define a method to link the pairs of points $(u_j, s_j), j=1, \dots, J$. An intuitive approximation would be to interpolate the observed income shares linearly. One major drawbacks of using linear interpolation is that these comparisons would be somewhat crude in that we assume that all the individuals classified in a given population group have the same income. Moreover, the Lorenz ordering is partial in the sense that not all distributions can be ranked. In these cases, we need to use inequality measures that provide a complete ranking of distributions. 
The Gini index has been the main indicator used to measure income inequality mainly due to its intuitive interpretation in terms of the area between the Lorenz curve and the egalitarian line. A nonparametric estimation of the Gini index is defined as twice the area between the egalitarian line and the Lorenz curve obtained by linear interpolation:

\begin{equation}\label{GiniE}
G(s_j,u_j) \approx \sum_{j=1}^J (s_j-s_{j-1})(u_j+u_{j-1}).
\end{equation}

The main limitation of computing the Gini index with the previous formula is that it yields biased estimates of inequality because its construction is based on the assumption that all individuals in a given population group get the same income. Hence, this formula is interpreted as the lower bound of the Gini coefficient, which neglects the variation within income shares (Cowell, \citeyear{cowell2011}). 

\begin{figure}\label{example}
\caption{Estimates of the Gini coefficient using different estimation techniques in Luxembourg and the Philippines}
\begin{tabular}{c c}
    \includegraphics*[scale=0.40]{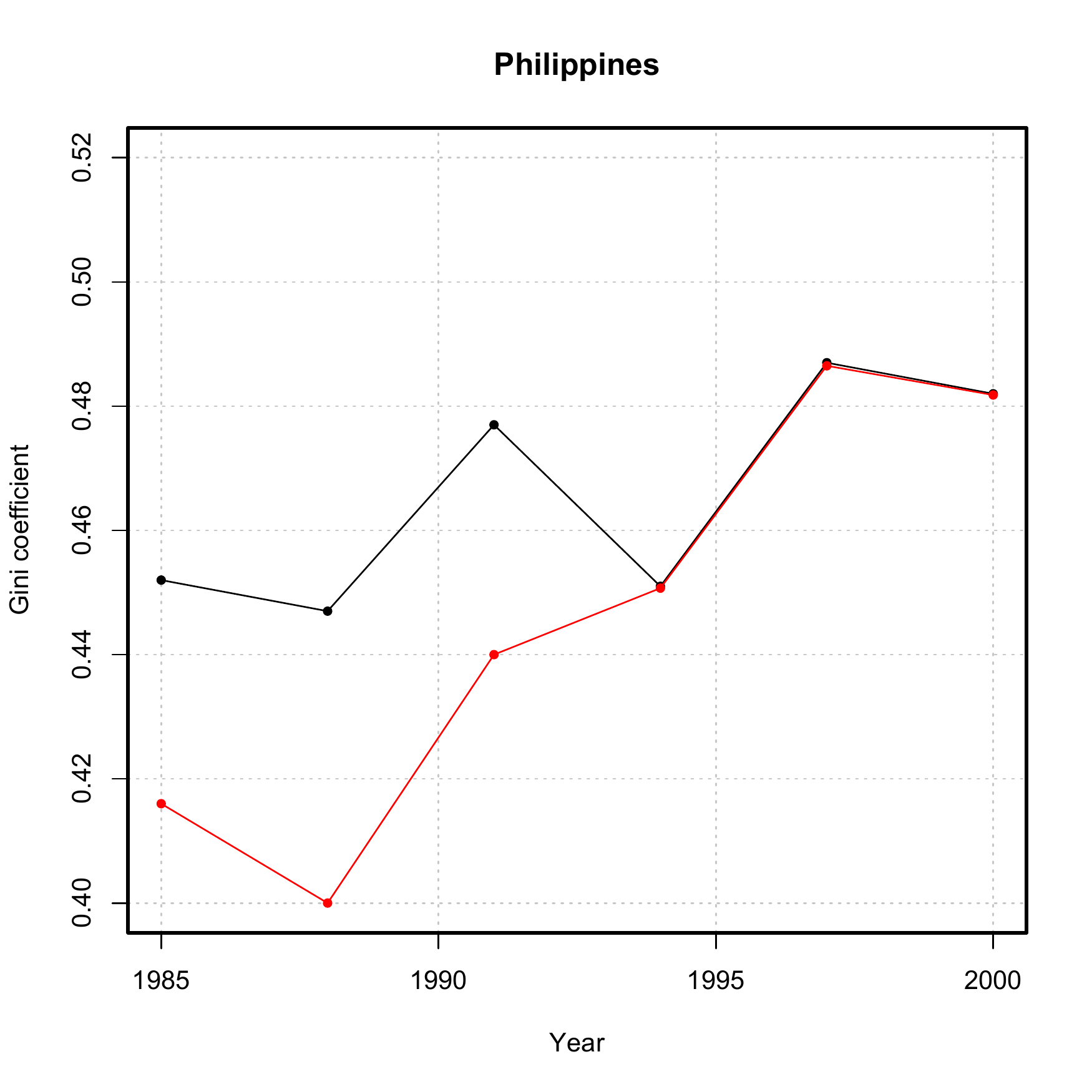}&\includegraphics*[scale=0.40]{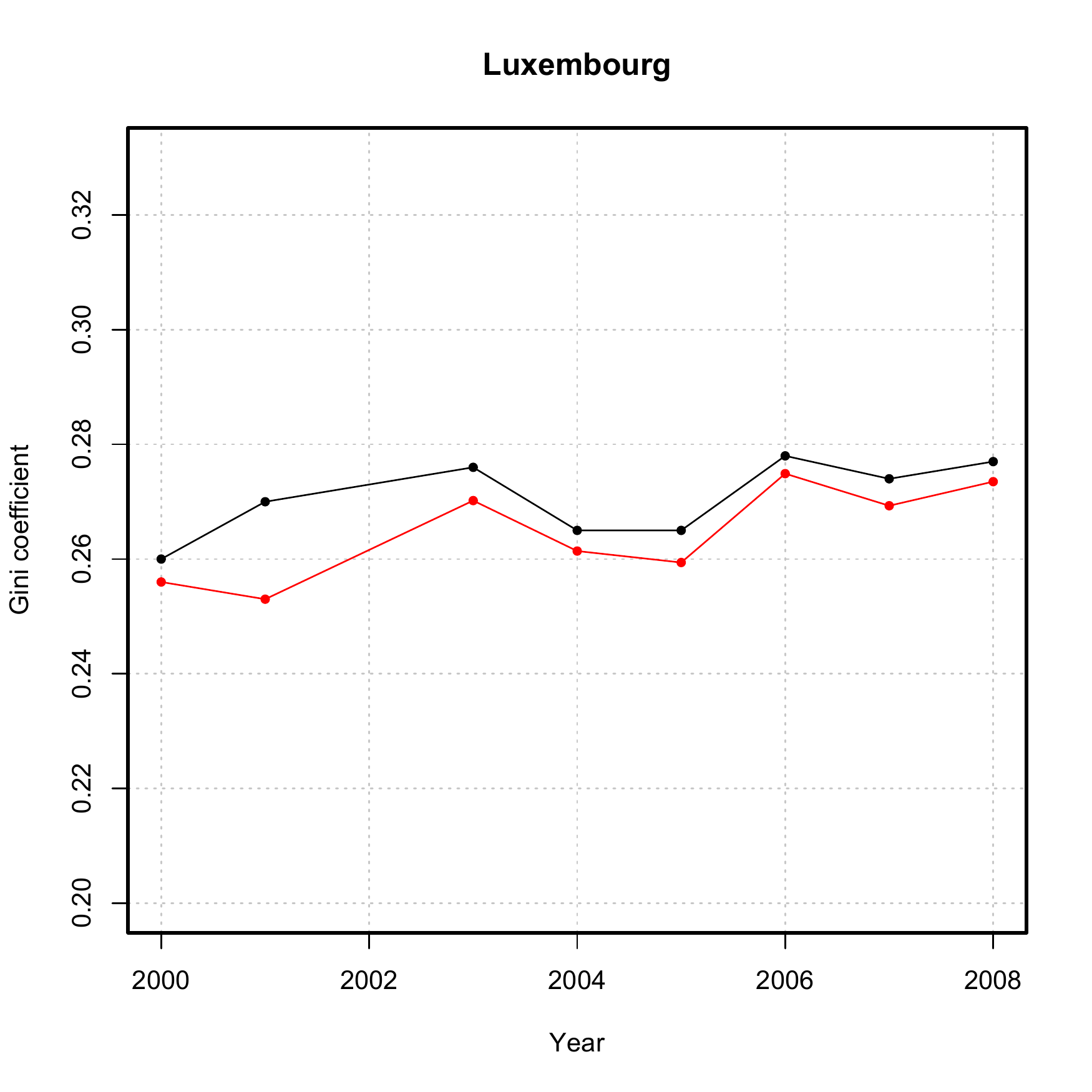}\\
 \end{tabular}

\footnotesize{Note: Black lines depict the evolution of the survey Gini index, red lines correspond to the evolution of lower bound of the Gini index .}

\end{figure}

Although this kind of analysis would yield biased estimates on inequality, it is expected to provide valuable information. Indeed, it is deemed to be useful because if an upward trend is observed, we could ensure that the overall disparities would also show an ascending pattern. Moreover, with optimal grouping, the bias is expected to be relatively small for observations with more than five data points (Davies and Shorrocks, \citeyear{davies1989}). Nevertheless, the empirical evidence based on this methodology is problematic in several ways. Firstly, the groups are not often optimally selected and, more importantly, the previous result has been obtained for the particular distribution of Canada. This finding might not necessarily match the distributional dynamics of other countries and, hence, the bias may be considerably higher than expected. Secondly, since the size of the bias might vary over time, it is not possible to obtain conclusions about the global evolution of income inequality, not even in those cases that exhibit an ascending trend. We illustrate this in Figure \ref{example}, which shows the evolution of the survey Gini coefficient in Luxembourg and the Philippines along with the lower bound of this measure, computed using Eq. (\ref{GiniE}). In the Philippines, the lower bound of the Gini index point towards an increase in income inequality from 1991 to 1994, but the survey Gini index show a downward trend during the same period. In Luxembourg, the survey Gini index rose one point from 2000 to 2001. The lower bound, however, fell from 0.256 to 0.253.

Parametric models are a sound statistical method to estimate inequality from grouped data. The use of a parametric model  aims at defining a more reliable approximation of the shape of the Lorenz curve between the observed income shares than a rough linear interpolation.\footnote{The nonparametric kernel density method has been also used to estimate income inequality (see, e.g. Sala-i-Martin, \citeyear{sala2006}). However, Minoiu and Reddy (\citeyear{minoiu2014}) already demonstrated the supremacy of the parametric models over the nonparametric techniques to estimate the income distribution. Hence, we do not focus on this statistical methodology in this paper.} However, it is key to chose a functional form that models accurately the income distribution. Among the whole range of alternatives, the GB2 family of distributions seems to be the most appealing option.\footnote{For a comprehensive review on this topic, we refer the reader to Kleiber and Kotz (\citeyear{kleiber2003}).}

 \subsection{The Generalized functions for the size distribution of income} \label{GB_family}

The generalized functions for the size distribution of income proposed by McDonald (\citeyear{Mcdonald1984}) include three well-known parametric models: the generalized beta of the first and the second kind (GB1 and GB2 respectively) and the generalized gamma (GG).  Among them, the GB2 distribution seems to be particularly suitable to model income distributions. It is a general class of distributions that is acknowledged to provide an accurate fit to income data (Jenkins, \citeyear{jenkins09}; McDonald and Xu, \citeyear{mcdonald1995}; McDonald and Mantrala,\citeyear{mcdonald1995m}). The GB2 can be defined in terms of the cumulative distribution function (cdf)  as follows:
\begin{equation*}
F(x;a,b,p,q)=B(v;p,q),\; x\ge 0,%
\end{equation*}
where $a,b,p,q>0$ and $v=(x/b)^a/[1+(x/b)^a]$. We represent by $B(v;p,q)=\int_0^ut^{p-1}(1-t)^{q-1}dt/B(p,q)$ the incomplete beta function ratio, where $B(p,q)=\int_0^1 t^{p-1}(1-t)^{q-1}dt$ is the beta function.

Let $\mathcal Z$ be the class of all non-negative random bariables with positive and finite expectation. For a random variable $X \in \mathcal Z$ with cdf $F(x; \boldsymbol\theta)$, we define $F^{-1}_X(y)= \inf \left\lbrace x; F_X(x)\geq y\right\rbrace$. The Lorenz curve associated with $X$ is defined as (Gastwirth, \citeyear{gastwirth1971})
\begin{equation}\label{LC2}
L_X(u) = \frac{\int_0^u F_X^{-1}(y)dy}{\int_0^1 F_x^{-1}(y) dy}, 0\leq u \leq 1.
\end{equation}%
Following Sarabia and Jorda (\citeyear{Sarabia2014}), the Lorenz curve in Eq. (\ref{LC2}) can also be expressed as,
\begin{equation}\label{LCgen}
L(u)=F_{X_{(1)}}(F^{-1}_{X}(u)), 0\leq u \leq 1,
\end{equation}%
where $F^{-1}_{Y}(u)$ denotes the quantile function and $F_{X_{(1)}}(x)=(1/E(X))\int_0^x tf(t)dt$ is the distribution of the first incomplete moment. To obtain the Lorenz curve we need, therefore, closed expressions for the cumulative distribution function and the distribution of the $k$th incomplete moment. These functions along with the $k$th moment and the Gini index are presented in Table \ref{GB2gini}.

Following Chotikapanich et al. (\citeyear{Chotikapanich2018}) and Arnold and Sarabia (\citeyear{arnold2018}), the Lorenz curve of the GB2 family is given by,
\begin{equation*}
L_{GB2}(u;a,p,q)=B\left(B^{-1}(u;p,q);p+\frac{1}{a},q-\frac{1}{a}\right),\;\;0\le u\le 1,
\end{equation*}
where $q>1/a$ and $B^{-1}(x;p,q)$ is the inverse of the incomplete beta function ratio.

This model nests most of the functional forms used to model income distributions including the beta of the second kind (Beta 2) when $a=1$, used by Chotikapanich et al. (\citeyear{chotikapanich12}) to estimate the global distribution of income; the Singh-Maddala (\citeyear{singh1976}) ($p=1$) and the Dagum (\citeyear{dagum1977}) ($q=1$) distributions, used by Hajargasht et al., (\citeyear{hajargasht2012}) and Bresson (\citeyear{bresson2009}). The Lorenz curves of these distributions can obtained using Eq. (\ref{LCgen}).
The Lorenz curve of the Second Kind Beta distribution can be expressed as follows:
\begin{equation}\label{LCB2}
L_{B2}(u;p,q)=B\left(B^{-1}(u;p,q);p+1,q-1\right),\;\;0\le u\le 1, q>1.
\end{equation}

\begin{landscape}
\begin{table}
\begin{center}

\caption{\label{GB2gini}
 Cumulative distribution function, $k$th moment distribution, $k$th moment and Gini index for a selection of distributions of the GB2 family}
\vspace{0.2cm}
 \small{
\begin{threeparttable}

\begin{tabular}{lcccc}
\toprule

Distribution & CDF & $k$th moment distribution & $E(X^k)$ & Gini Index   \\
\midrule
&&&&\\
GB2 & $\displaystyle B\left(\frac{(x/b)^a}{1+(x/b)^a};p,q\right)$ & $\displaystyle GB2\left(a,p+\frac{k}{a},q-\frac{k}{a}\right)$ & $\displaystyle \frac{b^kB(p+\frac{k}{a},q-\frac{k}{a})}{B(p,q)},\;q>k/a$ & see Eq. (\ref{giniGB2}) \\
&&&&\\
Second kind beta & $\displaystyle B\left(\frac{x/b}{1+x/b};p,q\right)$ & $\displaystyle B2(p+k,q-k)$ & $\displaystyle \frac{b^kB(p+k,q-k)}{B(p,q)},\;q>k$ &  $\displaystyle \frac{2B(2p,2q-1)}{pB^2(p,q)},\;\;q>1.$ \\
&&&&\\
Singh-Maddala  & $\displaystyle 1-\left(1+\left(\frac{x}{b}\right)^a\right)^{-q}$ & $\displaystyle GB2\left(a,1+\frac{k}{a},q-\frac{k}{a}\right)$ & $\displaystyle \frac{b^k\Gamma(1+\frac{k}{a})\Gamma(q-\frac{k}{a})}{\Gamma(q)},\;q>k/a$ & $\displaystyle 1-\frac{\Gamma(q)\Gamma(2q-\frac{1}{a})}{\Gamma(q-\frac{1}{a})\Gamma(2q)},\;\;q>1/a.$ \\
&&&&\\
Dagum & $\displaystyle \left(1+\left(\frac{x}{b}\right)^{-a}\right)^{-p}$ & $\displaystyle GB2\left(a,p+\frac{k}{a},1-\frac{k}{a}\right)$ & $\displaystyle \frac{b^k\Gamma(p+\frac{k}{a})\Gamma(1-\frac{k}{a})}{\Gamma(p)},\;k/a<1$ & $\displaystyle \frac{\Gamma(p)\Gamma(2p+\frac{1}{a})}{\Gamma(2p)\Gamma(p+\frac{1}{a})}-1,\;\;a>1.$ \\
&&&&\\
Lognormal & $\displaystyle \Phi\left(\frac{\log x-\mu}{\sigma}\right)$ & $\displaystyle LN(\mu+k\sigma^2,\sigma)$ & $\displaystyle \exp\left(k\mu+k^2\sigma^2/2\right)$ & $\displaystyle 2\Phi\left(\frac{\sigma}{\sqrt2}\right)-1.$ \\
&&&&\\
Fisk & $\displaystyle 1-\left(1+\left(\frac{x}{b}\right)^a\right)^{-1}$ & $\displaystyle GB2\left(a,1+\frac{k}{a},1-\frac{k}{a}\right),\;k/a<1$ & $\displaystyle b^k\Gamma(1+k)\Gamma(1-k),\;k<1$ & $\displaystyle \frac{1}{a},\;\;a>1.$ \\
&&&&\\
Weibull & $\displaystyle 1-e^{-(x/b)^a}$ & $\displaystyle GG\left(a,1+\frac{k}{a} \right)$ & $\displaystyle b^k\Gamma\left(1+\frac{k}{a}\right)$ & $\displaystyle 1-\frac{1}{2^{1/a}},\;\;a>1.$ \\
&&&&\\

\bottomrule
\end{tabular}

\begin{tablenotes}
\item Source: Arnold and Sarabia (\citeyear{arnold2018}), Kleiber and Kotz (\citeyear{kleiber2003}) and McDonald (\citeyear{Mcdonald1984}).
\item Note: $B(x;a,b)$ denotes the incomplete beta function. The existence of $k$th moment distribution, defined as $F_{(k)}(x)=(\int_o^x t^kdF(t))/(\int_0^{\infty}t^kdF(t)), x>0$, requires the same constraints about the parameters than the $k$th moment and $E(X^k) < \infty$.

\end{tablenotes}
\end{threeparttable}

}
\end{center}
\end{table}

\end{landscape}

For the Singh-Maddala distribution, the Lorenz curve is given by the following equation:
\begin{equation}\label{LCSM}
L_{SM}(u;a,q)=B\left(1-(1-u)^{1/q};1+\frac{1}{a},q-\frac{1}{a}\right),\;\;0\le u\le 1, q>1/a,
\end{equation}
and for the Dagum distribution it can be written as:
\begin{equation}\label{LCDA}
L_D(u;a,p)=B\left(u^{1/p};p+\frac{1}{a},1-\frac{1}{a}\right),\;\;0\le u\le 1, a>1.
\end{equation}

We also consider in this study two-parameter distributions, including the Fisk (\citeyear{fisk61}), which is a particular case of the GB2 making $p=q=1$, and the Weibull distribution, which is a special case of the GG distribution. We have also included the lognormal distribution as a limiting case of the GB2 distribution, which is one of the most popular candidates to model income variables (see e.g. Chotikapanich et al.,  \citeyear{chotikapanich1997}; Jorda et al., \citeyear{jorda2014}; Bresson \citeyear{bresson2009}). 

Using Eq. (\ref{LCgen}), we obtain the Lorenz curve of the lognormal distribution as:
$$L_{LN}(u;\sigma)=\Phi(\Phi^{-1}(u)-\sigma),\;\;0\le u\le 1,$$
where $\Phi(\cdot)$ represents the cdf of the standard normal distribution.
For the Fisk distribution, the Lorenz curve is given by
$$L_F(u;a)=B\left(u; 1+\frac{1}{a}, 1-\frac{1}{a}\right),\;\;0\le u\le 1,\;\;a>1. $$
Finally, for the Weibull distribution, the Lorenz curve is of the form
$$L_W(u;a)=G\left(-\log(1-u);\frac{1}{a}+1\right),\;\;0\le u\le 1,$$
where $G(x;\nu)=\int_0^x t^{\nu-1}\exp(-t)dt/\Gamma(\nu)$ is the incomplete gamma function ratio.

Closed expressions of the Gini index for some special cases of the GB2 family are summarized in Table \ref{GB2gini}. For the GB2 distribution, the Gini coefficient was provided by McDonald (\citeyear{Mcdonald1984}) and is given by,
\begin{equation}\label{giniGB2}
G_{GB2}=\frac{B(2q-1/a,2p+1/a)}{B(p,q)B(p+1/a,q-1/a)}\left(\frac{1}{p}J^{(1)}-\frac{1}{p+1/a}J^{(2)}\right),
\end{equation}
where
\begin{eqnarray*}
J^{(1)}&=&{}_3F_2\left(1,p+q,2p+\frac{1}{a};p+1,2(p+q);1\right),\\
J^{(2)}&=&{}_3F_2\left(1,p+q,2p+\frac{1}{a};p+\frac{1}{a}+1,2(p+q);1\right),
\end{eqnarray*}
if $q>1/a$, where ${}_3F_2(a_1,a_2,a_2;b_1,b_2;x)$ is a special case of the generalized hypergeometric function defined by
$$
\displaystyle {}_pF_q(a_1,\dots,a_p;b_1,\dots,b_q;x)=\sum_{k=0}^\infty\frac{(a_1)_k\cdots (a_p)_k}{(b_1)_k\cdots (b_q)_k}\frac{x^k}{k!},
$$
where $(a)_k$ represent the Pochhammer symbol defined by $(a)_k=a(a+1)\cdots(a+k-1)$.

\subsection{Estimation methods} \label{estimation}

To define a suitable estimation method, it is key to consider the manner in which the groupings have been generated. Hajargasht and Griffiths (\citeyear{Hajargasht2016}) recognise two different data generating process (DGP) that yield different methods for grouping observations. In the first process, the proportion of observations in each group is specified before sampling, so that the population proportions ($p_j$) are fixed, whereas income shares ($s_j$) are random variables. The second type of  DGP assumes pre-specified group boundaries $(g_j)$ and, hence, generates random population proportions in each interval. We focus on the first type of DGP because it fits the structure of the largest datasets of grouped data, including the WIID and PovcalNet. 

Traditionally, NLS has been used to estimate the vector of parameters of interest, regressing the income shares against the functional form of the Lorenz curve under the parametric assumptions made on the distribution of income. Let $X$ be a random variable in $ \mathcal Z$, with cdf $F(x; \boldsymbol\theta), \boldsymbol\theta \in \boldsymbol\Theta$ and Lorenz curve $L(u; \boldsymbol\theta)$, the estimation problem can be expressed as 
\begin{equation}\label{nls1}
\min_{\boldsymbol\theta}\sum_{j=1}^{J-1}(L(u_j; \boldsymbol\theta) - s_j)^2,
\end{equation}%
where $s_j$ is the income share held by the $j-th$ group and $u_j$ is its associated population proportion. The functional form of this curve for the distributions belonging to the GB2 family has been presented in Section \ref{GB_family}. 

As discussed before, the Lorenz curve is scale independent hence, using Eq. (\ref{nls1}), we are only able to estimate the subset of $\bm \theta$ corresponding to the shape parameters.\footnote{Hajargasht and Griffiths (\citeyear{Hajargasht2016}) proposed to use the generalised Lorenz curve to define the moment conditions. The generalised Lorenz curve is the result of scaling upward the ordinates of  Lorenz curve by mean income. With their approach, both scale and shape parameters can be estimated because the mean introduces the scale in the model.} The fact that we are only able to obtain estimates on shape parameters with this estimation procedure should not be interpreted as a limitation. Scale parameters are not needed to estimate relative inequality measures consistent with the Lorenz ordering, such as the Gini index or the Atkinson index. Therefore, if our interest resides in measuring relative inequality, this estimation strategy avoids the need to collect information on mean income. An additional advantage of this estimation strategy relative to the methods proposed in previous studies (see Hajargasht et al., \citeyear{hajargasht2012}) is that the income limits of the groups ($h_j$) are not estimated. Thus the dimensionality of the optimisation function is substantially reduced, which makes numerical optimisation simpler, especially when the number of moments is large (Chen, \citeyear{chen2017}).  

NLS estimation, however, overlook the fact that the sum of the income shares is, by definition, equal to one, thus introducing dependence between the income shares used in Eq. (\ref{nls1}). NLS yields, therefore, inefficient although still consistent estimates of $\bm \theta$ and hence of the functions that depend on this set of parameters, including relative inequality measures. To gain efficiency, we also deploy a GMM estimator of the following form:

\begin{equation}\label{GMM2}
\hat{\boldsymbol\theta}= \argmin_{\boldsymbol\theta} \bm{M}(\boldsymbol\theta)'\boldsymbol\Omega^{-1}\bm{M}(\boldsymbol\theta),
\end{equation}%
where $\bm{M}(\boldsymbol\theta)'= [m_1(\boldsymbol\theta), \dots, m_{J-1}(\boldsymbol\theta)]$ is the vector of moments conditions, which takes the form

\begin{equation}\label{estim}
\bm M(\boldsymbol\theta) = L(\bm u; \boldsymbol\theta) - \bm s,%
\end{equation}
being $\bm s' = (s_1, \dots, s_{J-1})$ a vector of cumulative income shares associated with the population proportions $\bm u' =(u_1, \dots, u_{J-1})$ and $\bm L(\bm u; \boldsymbol\theta)$ the theoretical Lorenz curve evaluated at $\bm u$. 

It should be noted that Eqs. (\ref{nls1}) and (\ref{GMM2}) are equivalent if $\bm \Omega= \bm I_{J-1}$. However, the identity matrix is not the optimal choice for $\bm \Omega$, which is why NLS yields less efficient estimates than GMM. The optimal choice of the weighting matrix $\bm \Omega$ is the variance and covariance matrix of the moment conditions. Results from Beach and Davison (\citeyear{beach1983}) and Hajargasht and Griffiths (\citeyear{Hajargasht2016}) characterise the asymptotic distribution of $\sqrt N(L(\bm u; \boldsymbol\theta) - \bm s)$ as a multivariate normal distribution with zero mean and variance and covariance matrix of the form:
\begin{equation*}\label{weightm}
\bm \Omega = \bm \Psi  \bm W \bm \Psi',%
\end{equation*}
where 
$$\bm \Psi = \left[ {\begin{array}{ccc:c}
   1/\mu &\dots &0 & -s_1/\mu\\
   \vdots  & \ddots & \vdots& \vdots  \\
    0 & \dots & 1/\mu & -s_{J-1}/\mu\\
  \end{array} } \right],$$
with $\mu=\int_0^{\infty}xf(x)dx.$ $\bm W$ is a symmetric matrix whose elements are 
$$\lbrace\bm W\rbrace_{i,j}= \mu_i^{(2)}+(u_ih_i-\mu s_i)(h_j-u_jh_j+\mu s_j)-h_i\mu s_i, \textrm{   for  } i\leq j$$%
with
$$\mu_i^{(2)}= \int_0^{g_i}x^2f(x)dx.$$

Because we only have access to grouped data on income shares, it is not possible to compute the variance and covariance matrix of the moment conditions. In order to obtain an efficient estimator of $\bm \theta$ from (\ref{GMM2}), we consider a two-step GMM estimator that uses the consistent estimates of NLS (Eq. (\ref{nls1})) to compute a first stage estimate of $\bm\Omega$, which is used in the second stage to estimate Eq. (\ref{GMM2}).

The estimation of Eq. (\ref{nls1}) involves the definition of starting values for the optimization algorithm.\footnote{We use the package \texttt{optim} in R to find the minimum of Eq. (\ref{estim}). BFGS algorithm is implemented by default and L-BFGS is used when this method reports an error. The computation of the gradient is done numerically.} For the two-parameter distributions presented, which only have one shape parameter, we propose to solve the following equation to obtain an initial value of $\theta$:
\begin{equation*}\label{GiniM}
g=G(\theta),%
\end{equation*}
 where $g$ is the sample Gini index, usually reported in the largest datasets of grouped income data, and $G(\theta)$ is the expression of the Gini index of the two-parameter distribution under consideration (see Table \ref{GB2gini}).

The distributions Singh-Maddala, Beta 2 and Dagum are characterised by two shape parameters, which complicates the definition of the non-arbitrary initial values. 
Conventionally, the estimates of a restricted model are  taken as initial values. A potential limitation of this method is that, as the dimensionality of the parameter space increases, it is more difficult to achieve global convergence. Although it seems quite intuitive that the moment estimates of the restricted model might be a good starting point, the optimization of the non-linear function in (\ref{nls1}) could converge to a local minima, which might lead to inaccurate estimates of the parameters and, hence, of inequality measures. The approach presented above for the two-parameter distributions is not feasible for these models in most cases because no other information besides the Gini index and the income shares is reported. To provide several non-arbitrary combinations of starting values, we propose to use the following procedure:
\begin{enumerate}
\item Rewrite Eq. (\ref{nls1}) using the Lorenz curve of the model to be estimated $L(u;\theta_1,\theta_2)$, which is given in Eq. (\ref{LCB2}) for the B2 distribution, in Eq. (\ref{LCSM}) for the Singh-Maddala distribution and in Eq. (\ref{LCDA}) for the Dagum distribution.
\item Define a  grid of integer numbers for the starting values of  $\theta_1$, $\theta_1^{(s)}\in [1,20].$
\item Solve $g=G(\theta_1^{(s)},\theta_2)$ for $\theta_2$, to obtain $\theta_2^{(s)}$.
\item Estimate the Eq. (\ref{nls1}) using $(\theta_1^{(s)},\theta_2^{(s)})$, as initial values.
\item Keep the parameter estimates with the lowest residual sum of squares (RSS).
\end{enumerate}

The routine described above allows us to obtain the moment estimates of one of the parameters assuming that the other one is equal to the grid value. These 20 combinations of initial values are used to undertake 20 different regressions using Eq. (\ref{nls1}). Although we cannot ensure that our estimates belong to the global minima our proposed procedure covers a larger proportion of the parametric space than just using the moment estimates of a particular sub-model. 

For the GB2 distribution, which has three shape parameters and one scale parameter, we make use of the estimates obtained for the three-parameter distributions. We perform the estimation of Eq. (\ref{nls1}) using as initial values the 20 combinations of parameters from the Beta 2 distribution, setting $a=1$; the 20 initial values of the Singh-Maddala distribution with $p=1$; and the ones obtained for the Dagum distribution, assuming $q=1$. We save the estimation that reports the best fit.

For the estimation $\bm\Omega$, we also need to compute the mean ($\mu$), the second order moment ($\mu_j^{(2)}$) and the income limits of each group($h_{j}$). Therefore, even though we are interested in scale independent inequality measures, we need to estimate consistently the scale parameter. Let $\eta$ denote the scale parameter of the distribution so that, $\bm \theta = (\eta, \bm\lambda)' $. We propose to estimate $\eta$ by solving the following equation:
\begin{equation}\label{media}
\bar X = \mu(\eta, \bm\lambda),%
\end{equation}
where $\bar X$ is the sample mean and $\mu(\eta, \bm\lambda) = \int_{\R_+}xf(x; \bm\theta)dx$, whose expressions for the distributions belonging to the GB2 family are presented in Table \ref{GB2gini}.

Let $\theta^* = (\eta^*, \bm\lambda^*)'$  be the consistent estimator of the parameters of the model obtained from Eqs.(\ref{nls1}) and (\ref{media}) used to obtain a first-stage estimator of the weighting matrix ($\bm\Omega^*=\bm\Omega(\theta^*)$).  Substituting $\bm\Omega^*$ in (\ref{GMM2}), we obtain the second-stage estimator of $\bm\lambda$ as:
\begin{equation}\label{GMM3}
\hat{\boldsymbol\theta}= \argmin_{\boldsymbol\theta} \bm{M}(\boldsymbol\theta)'\boldsymbol\Omega^{*-1}\bm{M}(\boldsymbol\theta).
\end{equation}

Because the first stage estimator ($\Lambda^*$) from Eq. (\ref{nls1}) is consistent, so is the weighting matrix $\boldsymbol\Omega^*$. The replacement of $\bm\Omega$ by a consistent estimate of this matrix does not affect to the asymptotic properties of the GMM estimator. It affects, however, to the small-sample behavior of this estimator, which is generally biased (Altonji and Segal, \citeyear{Altjoni1996}). Prior research based on Monte Carlo simulation suggest that the size of the bias depends on the underlying distribution of the data, being particularly large for heavy-tailed distributions (Altonji and Segal, \citeyear{Altjoni1996}). The size of the bias increases with the number of overidentifying restrictions (Clark, \citeyear{clark1996}). Although this limitation is overcome as the sample size gets large (Hansen, \citeyear{hansen1982}), information on the sample size used to construct the grouped data is not often available. It is, therefore, recommended to deploy both estimators and, if the parameter estimates differ substantially, opt for NLS results.

\section{Results}
\subsection{Estimation of the Gini coefficient using grouped data} \label{macro}

In this section, we explore some practical issues in the estimation of economic inequality from grouped data. To consider a diverse set of observations, we use the most comprehensive and up-to-date source of grouped income data: the latest version WIID, released in January 2017. This database collects information for more than  8817 datasets for 182 countries over the period 1867-2015. Each dataset might report different types of information: the Gini index is generally provided ($99.6\%$ of the observations); less frequently we can find information on five to ten income shares ($63.2\%$); finally, it is less common to report data on mean income ($50.1\%$). 

Before going any further, it should be mentioned that the WIID3.4 stores a heterogeneous collection of datasets in terms of welfare concept, unit of analysis, equivalence scale, quality of the data and population and area coverage. For this reason, the WIID also includes additional data about these concepts along with information on the source from which the data was taken. Therefore, even though the WIID is notable in terms of their geographical and time coverage, the lack fo data comparability is often recognised as a potential limitation. In this study, however, we take advantage of such heterogeneity to illustrate whether these data issues affect to the performance of the different estimation methods in order to bring them to the attention of potential users.

The first question that arises when deploying the parametric approach is: given that we expect that grouped data collected by the WIID comes from surveys of reasonable sample size, is the asymptotically efficient GMM a superior method than the unbiased NLS to estimate income inequality? To answer this question we estimate the different parametric distributions belonging to the GB2 family using both econometric strategies. For the two sets of parameter estimates obtained from Eqs. (\ref{nls1}) and (\ref{GMM3}), we compute the Gini coefficient using the expressions presented in Table \ref{GB2gini}. These two estimates are compared with the observed Gini index reported by the survey. Because our interest resides in measuring income inequality, we opt for the method that yields more accurate estimates of the Gini coefficient.\footnote{The fact that one method provides more accurate estimates of the Gini index does not mean that it is the most suitable econometric strategy to model relative inequality. To provide strong evidence in this regard, we should examine the robustness of this result to the consideration of different inequality measures. Unfortunately, other measures besides the Gini coefficient are rarely reported. However, we further explore this issue in Section \ref{roboust} with individual records.}

Since the optimisation function in Eq. (\ref{nls1}) depends only on shape parameters, NLS can be deployed in the 5570 country/year datasets which present information on, at least, 5 income shares. Because the Gini coefficient is scale-independent, the estimation of the shape parameters is enough to estimate this inequality measure. To obtain the GMM estimates in Eq. (\ref{GMM3}), we need besides data on income shares, information about the mean of the income distribution to estimate the scale parameter, which is used to construct a consistent estimate of $\bm\Omega$ (Eq.\ref{weightm}). As a result, this method can be implemented only in 3286 country/year datasets of the WIID. Table \ref{tablecomp} presents the proportion of observations for which GMM dominates NLS in the estimation of the Gini index. 

Our estimates for the GB2 distribution suggest that GMM reports more accurate estimates than NLS only for 25\% of the datasets. This proportion tends to decrease with the number of parameters of the distributions. On the contrary, we observe that the GMM estimator presents higher dominance rates when fewer income shares are considered. Therefore, in line with the findings presented by different simulation studies, these two results confirm that the size of the bias increases with the number of overidentifying restrictions (Clark, \citeyear{clark1996}). In sum, our results suggest that NLS is generally preferred over GMM to estimate the Gini coefficient. Hence, researchers may be willing to sacrifice asymptotic efficiency if the priority is placed on obtaining unbiased parameter estimates. In that case, NLS would be preferred over GMM, even in situations where GMM is by far superior in asymptotic efficiency (Aljoni and Segal, \citeyear{Altjoni1996}).

\begin{table}
\begin{center}

\caption{\label{tablecomp}
 Comparison of the performance of the GMM and the NLS estimators of the Gini coefficient}
\vspace{0.2cm}
 \small{
\begin{threeparttable}

\begin{tabular}{l c c c c c c c}
\toprule

				&	GB2	&	Beta 2	&	Singh-Maddala	&	Dagum	&	Lognormal	&	Fisk	&	Weibull	\\
	\midrule
Total	 			&	25.26\%	&	24.33\%	&	14.87\%	&	8.99\%	&	12.05\%	&	10.91\%	&	0.76\%	\\
5 income shares	&	37.66\%	&	52.50\%	&	34.86\%	&	32.46\%	&	23.31\%	&	25.65\%	&	3.06\%	\\
10 income shares	&	25.66\%	&	22.93\%	&	13.84\%	&	7.73\%	&	11.44\%	&	10.01\%	&	0.66\%	\\\bottomrule
\end{tabular}

\begin{tablenotes}
\item Note: Results based on the 3286 datasets of the WIID, of which 154 presented information on 5 income shares and 3132 data on 10 income shares. For all parametric distributions except the GB2, the Gini coefficient has been estimated using the formulas in Table \ref{GB2gini}. The Gini index of the GB2 distribution has been estimated by Monte Carlo simulation using samples of size $N=10^6$.

\end{tablenotes}
\end{threeparttable}

}
\end{center}
\end{table}

The second question of interest when examining inequality from grouped data is whether parametric functional forms provide better approximations of the Lorenz curve than the nonparametric method of linear interpolation. The relevance of this question resides in the overwhelming number of studies that have opted for linear interpolation to construct the Lorenz curves and its associated inequality measures (see Anand and Segal (\citeyear{anand2008}) for a review). The  popularity of this approach seems to be explained by its simplicity and supported by the extended argument that parametric functional forms might lead to misspecification bias because their implementation requires making \textit{ex-ante} assumptions about the shape of the income distribution and/or the Lorenz curve. To compare the performance of the GB2 distribution and the related sub-models to estimate income inequality with the nonparametric approach, conventional measures of goodness of fit (GOF), such as the residual sum of squares, are not informative because linear interpolation is designed to perfectly match the income shares. Hence, as a measure of GOF, we consider the gap between the survey Gini coefficient and the estimated Gini indices using both the nonparametric lower bound (Eq.\ref{GiniE}) and the different parametric functional forms (Table \ref{GB2gini}). 

For the estimation of the parametric models, we focus on NLS estimates because this method seems to yield more accurate estimates of the Gini index.\footnote{We present the results for the GMM estimates in the Appendix, which seem to confirm the outperformance of the parametric models in general, and the GB2 distribution in particular.} An additional advantage of analysing NLS estimates is that it allows us to examine a larger number of observations. We have reformulated Eq. (\ref{nls1}) to estimate seven parametric distributions that belong to the GB2 family for the 5570 country/year datasets with at least five income shares available. The estimated Gini indices have been computed using the closed formulas presented in Table \ref{GB2gini} except for the GB2 distribution, for which the Gini coefficient has been estimated by Monte Carlo simulation. We have also calculated the lower bound of the Gini index derived from linear interpolation of the Lorenz curve. This approximation of the Gini index assumes equality of incomes within shares. Hence, its value must be lower than the survey Gini index computed with individual records because it considers the existing variation within income shares. We found, however, that this relation was violated in 355 datasets. This incongruent result might have two potential explanations. Because Eq. (\ref{GiniE}) is an approximation of the lower bound, it has an inherent error that may lead to such inconsistencies. It could be also explained by measurement errors in the  Gini index or the income shares included in the WIID database. Hence, we opt for removing those cases to facilitate the discussion of the results.\footnote{Overall, the estimates for the whole sample with 5570 country/year datasets point out the same conclusions as for the restricted sample (results available upon request).}

Table \ref{DiffNLS} presents the difference between the survey Gini index and the parametric and nonparametric estimates. To facilitate the comparison of these two methodologies, we report the results in absolute value. Our results reveal that the lower bound yields a very poor approximation of the Gini index. The gap with the observed Gini index is above 0.01 in 56\% of the cases. The parametric approach, instead, provides much more accurate results, with substantially lower differences between the estimated and the observed Gini index. On average, lower bound estimates report an error three to four times larger than most parametric models. Among the parametric models, the GB2 seems to outperform the other sub-models, with estimation errors lower than 0.01 for 92\% of observations. As regards the particular cases of this family, even the two-parameter distributions report fairly accurate Gini indices, which differ in less than 0.02 in 95\% of the cases. Estimations of  the Gini index with errors larger than 0.1 are more frequent for the nonparametric approach. All parametric specifications report the same proportion of estimates with differences larger than 0.1, the 0.6\%, corresponding to three datasets: Mauritius in 1980, Zambia in 2004, rural and urban. In these three cases, the parametric and the nonparametric approaches report very similar estimates of the Gini index. For instance, in rural Zambia the lower bound is 0.189 and the estimate for the GB2 distribution is 0.188, but the WIID reports a survey Gini coefficient of 0.55. Hence, we believe, that the survey data of these datasets may present some kind of measurement error.

\begin{table}
\begin{center}

\caption{\label{DiffNLS}
Absolute error in the estimation of the Gini index using linear interpolation and different parametric distributions of the GB2 family}
\vspace{0.2cm}
 \small{
\begin{threeparttable}

\begin{tabular}{l c c c c c c}
\toprule
Distribution	&	Mean	&	[0, 0.01)	&	[0.01, 0.02)	&	[0.02, 0.05)	&	[0.05, 0.1)	&	[0.1, )	\\
\midrule
Lower bound	&	0.0140	&	44.28\%	&	39.5\%	&	14.57\%	&	1.42\%	&	0.23\%	\\
GB2	&	0.0034	&	92.20\%	&	4.60\%	&	2.36\%	&	0.79\%	&	0.06\%	\\
B2	&	0.0040	&	91.68\%	&	4.99\%	&	2.53\%	&	0.75\%	&	0.06\%	\\
SM	&	0.0040	&	91.47\%	&	5.25\%	&	2.49\%	&	0.73\%	&	0.06\%	\\
Dagum	&	0.0041	&	91.24\%	&	5.48\%	&	2.49\%	&	0.73\%	&	0.06\%	\\
Log-normal	&	0.0043	&	91.26\%	&	5.64\%	&	2.24\%	&	0.81\%	&	0.06\%	\\
Fisk	&	0.0043	&	91.18\%	&	5.45\%	&	2.61\%	&	0.71\%	&	0.06\%	\\
Weibull	&	0.0062	&	85.64\%	&	10.78\%	&	2.82\%	&	0.71\%	&	0.06\%	\\
\bottomrule
\end{tabular}

\begin{tablenotes}
\item Note: Results based on 5215 datasets of the WIID. Parametric models have been estimated by NLS. The lower bound of the Gini index has been obtained using Eq. (\ref{GiniE}). For the parametric distributions except the GB2, the Gini coefficient has been estimated using the formulas in Table \ref{GB2gini}. The Gini index of the GB2 distribution has been estimated by Monte Carlo simulation using samples of size $N=10^6$.
\end{tablenotes}
\end{threeparttable}
}
\end{center}
\end{table}

Although the previous results point towards a better performance of the parametric models over the nonparametric approach, only 16\% of the observations show large deviances (higher than 0.05) between the lower bound and the observed Gini index. Therefore, it could be argued that the nonparametric approach provides the researchers with an intuitive and fairly accurate tool to assess inequality in most cases. However, since our sample includes Gini coefficients of very different magnitude, the error should be evaluated in relative terms. Table \ref{RDiffNLS} shows the absolute difference between the observed and the estimated Gini coefficient, relative to the value reported by survey. These results strongly suggest that the lower bound yields considerably inaccurate estimates of the Gini index, which is underestimated by more than 2\% in the 73\% of the country/year observations of the WIID. 

These estimates reflect not only that linear interpolation is a poor approximation of the Lorenz curve, but
also that parametric distributions lead to highly reliable estimates of the Gini index. The GB2 distribution seems to offer the best estimates, with 84\% of estimations providing Gini coefficients that deviate less than 1\% from the survey Gini index; for the three-parameter functional forms, this proportion goes down to 80\%. The two-parameter functional forms also present fairly accurate results for the 70\% of observations, except for the Weibull distribution.

\begin{table}
\begin{center}

\caption{\label{RDiffNLS}
Relative error in the estimation of the Gini index using linear interpolation and different parametric distributions of the GB2 family}
\vspace{0.2cm}
 \small{
\begin{threeparttable}

\begin{tabular}{l c c c c c}
\toprule
Distribution	&	[0\%, 1\%)	&	[1\%, 2\%)	&	[2\%, 5\%)	&	[5\%, 10\%)	&	[10\%, )	\\
         \midrule
Lower bound	  	&	3.16\%	&	10.11\%	&	72.66\%	&	11.01\%	&	3.07\%	\\
GB2			&	84.7\%	&	5.66\%	&	6.27\%	&	2.15\%	&	1.23\%	\\
B2			&	77.01\%	&	12.54\%	&	6.98\%	&	2.24\%	&	1.23\%	\\
SM			&	80.36\%	&	9.11\%	&	7.11\%	&	2.19\%	&	1.23\%	\\
Dagum		&	80.19\%	&	9.03\%	&	7.5\%	&	2.11\%	&	1.17\%	\\
Lognormal	&	72.58\%	&	16.03\%	&	7.96\%	&	2.21\%	&	1.23\%	\\
Fisk			&	74.96\%	&	14.06\%	&	7.54\%	&	2.26\%	&	1.19\%	\\
Weibull		&	48.9\%	&	35.92\%	&	11.95\%	&	2.11\%	&	1.13\%	\\\bottomrule
\end{tabular}

\begin{tablenotes}
\item Note: Results based on 5215 datasets of the WIID. Parametric models have been estimated by NLS. The lower bound of the Gini index has been obtained using Eq. (\ref{GiniE}). For the parametric distributions except the GB2, the Gini coefficient has been estimated using the formulas in Table \ref{GB2gini}. The Gini index of the GB2 distribution has been estimated by Monte Carlo simulation using samples of size $N=10^6$.

\end{tablenotes}
\end{threeparttable}

}
\end{center}
\end{table}

Due to the inherent heterogeneity of the WIID in terms of welfare definition and data quality, another relevant question is whether these data characteristics affect accuracy of the previous estimates.  More importantly, does the estimation error decrease with the number of income shares? The answer to the last question is quite obvious for the nonparametric approach: the larger the number of income shares, the better the approximation of the Lorenz curve, hence the more reliable the estimate of the Gini coefficient. For parametric models, however, five income shares might be enough to represent the shape of the Lorenz curve as rigorously as with 10 data points.

Table \ref{table2} presents a summary of the absolute error in the estimation of the Gini index using the GB2 distribution, as the best parametric approximation of the Lorenz curve, and linear interpolation. We present the mean and the standard deviation (in parenthesis) of the difference in absolute terms between the survey and the estimated Gini coefficient for the four new welfare categories introduced in the last version of the WIID: consumption, disposable income, gross income and others;\footnote{This new classification is a simplified version of the previous classification by welfare definition, which combines categories that are close to each other. See \url{https://www.wider.unu.edu/sites/default/files/Data/WIID3.4} for a detailed description of the new labels.} and for different data quality levels: high, average, low and not known. To examine the effect of using a larger number of moments, we present these results disaggregated by five and ten income shares. In this regard, our results suggest that, using linear interpolation, the error in the estimation of the Gini with only five income shares index is twice to three times higher than in datasets with ten data points. In the parametric framework, this pattern is not so obvious. Overall, we find the estimates performed with a larger number of income shares lead to more accurate estimates. The difference in the estimation error might be considerable in some categories, such as \textit{disposable income} or \textit{high data quality}, with estimation errors four to five times larger when five income shares are used. In other categories, however, the accuracy of the estimates does not seem to be affected by the number of moments. Low-quality datasets show, on average, estimation errors of the same magnitude, but with higher variation for the estimation with ten data points. Hence, we might find larger estimation errors in datasets with ten income points than with five, for this particular category. This result should not be interpreted as a recommendation to use five income shares for the estimation of parametric models with datasets of poor quality. Instead, this should be seen as an argument in favor of the parametric approach, which even with very few points of the Lorenz curve might yield reliable estimates.

\begin{table}
\begin{center}

\caption{\label{table2}
Absolute error in the estimation of the Gini index using the GB2 distribution and linear interpolation}
\vspace{0.2cm}
 \small{
\begin{threeparttable}

\begin{tabular}{l l c c c c}
\toprule
	& & 	\multicolumn{2}{c}{GB2 distribution}			&	\multicolumn{2}{c}{Linear interpolation}	\\
\midrule
	&		&	10 shares	&	5 shares	&	10 shares	&	5 shares	\\
\midrule
	&	Consumption        		&	0.0048	&	0.0158	&	0.0111	&	0.0292	\\
	&	(857, 124)				&	(0.0139)	&	(0.0330)	&	(0.0097)	&	(0.0264)	\\
Welfare	&	Income, disposable	&	0.0032	&	0.0118	&	0.0124	&	0.0312	\\
definition	&	(2686, 112)		&	(0.0081)	&	(0.0185)	&	(0.0085)	&	(0.0206)	\\
	&	Income, gross			&	0.0110	&	0.0125	&	0.0116	&	0.0272	\\
	&	(384, 325)				&	(0.0155)	&	(0.0153)	&	(0.0120)	&	(0.0162)	\\
	&	Other				&	0.0031	&	0.0082	&	0.0132	&	0.0329	\\
	&	(1031, 51)				&	(0.0147)	&	(0.0152)	&	(0.0157)	&	(0.0158)	\\
\midrule
	&	Average				&	0.0031	&	0.0076	&	0.0108	&	0.0268	\\
	&	(1645, 139)			&	(0.009)	&	(0.0136)	&	(0.0076)	&	(0.0159)	\\
Data 	&	High				&	0.0026	&	0.0135	&	0.0122	&	0.0324	\\
quality	&	(2559, 167)		&	(0.0079)	&	(0.0298)	&	(0.0086)	&	(0.0243)	\\
	&	Low					&	0.0125	&	0.0130	&	0.0159	&	0.0265	\\
	&	(642, 261)				&	(0.0226)	&	(0.0161)	&	(0.0210)	&	(0.0167)	\\
	&	Not known			&	0.0022	&	0.0236	&	0.0131	&	0.0352	\\
	&	(112, 45)				&	(0.0025)	&	(0.0164)	&	(0.0076)	&	(0.0224)	\\	\\
\bottomrule
\end{tabular}

\begin{tablenotes}
\item Note: The number of datasets used to compute the mean and the standard deviation of the error in the estimation of the Gini index are presented in parenthesis below the label of the corresponding category, for ten and five income shares respectively. The GB2 distribution has been estimated by NLS and the Gini index has been computed by  Monte Carlo simulation using samples of size $N=10^6$.

\end{tablenotes}
\end{threeparttable}

}
\end{center}
\end{table}

Despite the fact that we are primarily interested in assessing income inequality, only comparing  Gini indices we are not able to assert the supremacy of any parametric model. To provide a complete picture of the GOF of the different parametric models, we turn now our attention to measures that evaluate the performance of nested models considering not only the accuracy, but also the parsimony of the model by penalizing for the number of parameters.\footnote{The results for the Schwartz Bayesian Information Criteria and weighted sum of the residuals can be found in the Appendix, Tables \ref{BIC table} and \ref{Wtable} respectively.} Table \ref{AIC table} presents the proportion of observations for which the models in rows outperform the distributions in columns according to the Akaike Information Criterion (AIC). Our results suggest again the GB2 distribution is a more suitable model for income and consumption variables, although the three-parameter models seem to be preferred in about 15 percent of the cases when penalizing by the number of parameters. As regards the three-parameter distributions, the Beta 2 and the Dagum distributions seem to perform equally well, but the Singh-Maddala distribution seems to yield more accurate estimates in most cases. As expected, the two-parameter models rarely improve the GOF of the GB2 and the three-parameter functional forms. These figures show similar dominance rates are observed for the lognormal and the Fisk distributions and confirm that the Weibull distribution is the least suitable model to represent income distributions.
 
\begin{table}
\begin{center}

\caption{\label{AIC table}
GOF dominance matrix based on the AIC}
\vspace{0.2cm}
 \small{
\begin{threeparttable}

\begin{tabular}{l c c c c c c c}
\toprule
			&	GB2	&	Beta 2	&	Singh-Maddala	&	Dagum	&	Lognormal	&	Fisk	&	Weibull	\\
	\midrule
GB2	  		&	100\%	&	87\%	&	85\%	&	88\%	&	98\%	&	98\%	&	99\%	\\
Beta 2		&	13\%		&	100\%	&	43\%	&	48\%	&	96\%	&	84\%	&	98\%	\\
Singh-Maddala	&	15\%		&	57\%	&	100\%	&	62\%	&	93\%	&	89\%	&	99\%	\\
Dagum		&	12\%		&	52\%	&	38\%	&	100\%	&	87\%	&	89\%	&	99\%	\\
Lognormal	&	2\%		&	4\%	&	7\%	&	13\%	&	100\%	&	45\%	&	95\%	\\
Fisk			&	2\%		&	16\%	&	11\%	&	11\%	&	55\%	&	100\%	&	91\%	\\
Weibull		&	1\%		&	2\%	&	1\%	&	1\%	&	5\%	&	9\%	&	100\%	\\
\bottomrule
\end{tabular}

\begin{tablenotes}
\item Note: Results based on 5215 datasets of the WIID. Parametric models have been estimated by NLS.
\end{tablenotes}
\end{threeparttable}

}
\end{center}
\end{table}

These figures corroborate the results of previous studies (McDonald, \citeyear{Mcdonald1984}; Butler and McDonald, \citeyear{butler1986}; Hajargasht et al., \citeyear{hajargasht2012}) about the excellent performance of the GB2 distribution to describe income data. To our knowledge, this is the first study that analyzes the suitability of this functional form and its related sub-models for a large number of countries and for such a long period of time, thus generalizing the result as regards the flexibility of this family to represent income data for a heterogeneous sample of countries. 

\subsection{Simulation of grouped data from individual records: a robustness check}\label{roboust}

\begin{figure}
\caption{\label{incosn}
Lorenz curve of Argentina (1961): Singh-Maddala (red) and lognormal (black) distributions}
\begin{center}
\includegraphics*[scale=0.47]{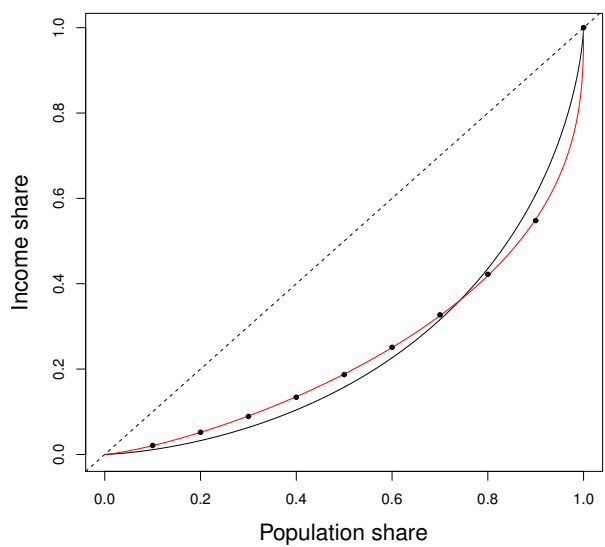}
\end{center}
\end{figure}

So far, our analysis suggests that the GB2 family includes excellent models to obtain reliable estimates of the Gini index. However, the analysis of income inequality rarely relies on just one measure. Depending on the properties of the inequality measures and their sensitivity to different parts of the distribution inequality indices may reflect different evolutions. Hence, the evaluation of the performance of different models to estimate income inequality should not be based solely on the Gini coefficient. Because this index is constructed as a function of the area between the Lorenz curve and the egalitarian line, differences below the observed income shares can be offset by overestimated income shares. This case is illustrated in Figure \ref{incosn}, which presents the survey income shares (black points) of Argentina in 1961 and the fits of the Singh-Maddala (red line) and the lognormal (black line) distributions. This graph reveals that the Singh-Maddala distribution provides a highly accurate fit and clearly outperforms the lognormal distribution. However, a comparison of the survey  and the estimated Gini coefficients suggests a better performance of the lognormal distribution to estimate income inequality: the survey Gini index is 0.531 and the estimated Gini indices of the Singh-Maddala  and the lognormal distributions are 0.516 and and the 0.522 respectively. 

The apparent outperformance of the lognormal distribution is, therefore, a statistical artifact caused by the manner in which the Gini coefficient is defined. If we compared the observed and the estimated values of inequality measures more sensitive to the left tail, the Singh-Maddala distribution would be declared as a superior model. For this kind of measures, the lognormal distribution would overestimate inequality levels because its Lorenz curve of lies far below the sample income shares at the bottom of the distribution. Unfortunately, other measures besides the Gini coefficient are not reported in the WIID. 

To extend the insights about the estimation of income inequality from grouped data, we rely on data from the Luxembourg Income Study (LIS). The LIS database gathers harmonised microdata on disposable income collected from nearly 50 countries, which cover the period from 1980 to 2016. Using the 278 datasets of individual records available in the ten waves of the LIS database, we reconstruct grouped data with the same structure as the WIID: five and ten income shares, the mean and the Gini coefficient. These statistics have been obtained following the methodological guidelines of LIS.\footnote{A detailed description of these guidelines can be obtained from \url{http://www.lisdatacenter.org/data-access/key-figures/methods/} and the R code used for the computation of inequality measures can be downloaded from \url{http://www.lisdatacenter.org/wp-content/uploads/files/access-key-programs-r-ineq.txt}} We consider equivalised disposable income which is equal to household income divided by the square root of the household size. We exclude all missing observations and records with zero disposable income. To the remaining sample, LIS proposes to apply top and bottom coding. Equivalised income is bottom-coded at 1\% of equivalised mean and top-coded at 10 times the median household income.\footnote{The aim of this section is to expand the results presented in Section \ref{macro}. To do so, it is essential to replicate as accurately as possible the context of limited information under which those results were obtained.  Hence, although bottom and top coding applied the income variable might be debatable, we apply this procedure not only because of LIS recommendations but because WIID data from LIS is reported with censoring. For this reason, even though we are deliberately introducing some kind of measurement error, we do not consider the double censoring in the estimation of the parametric models because we do not have this kind of information when using grouped data.} Finally, household weights are multiplied by the household size to obtain person-adjusted weights.

The advantage of working now with individual data is that we do not have to restrict the analysis to the Gini coefficient. To examine the reliability of the parametric models to estimate different inequality measures, we have also calculated the Atkinson index of the surveys using the following expression:
\begin{equation*}
A_{\epsilon}=1 - \left(\frac{1}{N}\sum_{i=1}^N \left(\frac{x_i}{\mu} \right)^{1-\epsilon} \right)^{\frac{1}{1-\epsilon}} , \epsilon\neq 1,
\end{equation*}%
\begin{equation*}
A_{\epsilon}=1 - \frac{1}{\mu}\prod_{i=1}^N x_i^{1/N}  , \epsilon= 1,
\end{equation*}%
where $\epsilon$ is an inequality aversion parameter, which makes this measure more sensitive to the left tail of the income distribution as it increases.  

The income shares from LIS are used to replicate the analysis of Section \ref{macro}, comparing estimated and survey inequality measures. Tables \ref{micro10} and \ref{micro5} present the absolute error in the estimation of the Gini coefficient and the Atkinson indices from grouped data in form of 10 and 5 income shares respectively. The parameters of the parametric models have been estimated by NLS following the procedure presented in Section \ref{estimation}.\footnote{Results based on the estimation of parametric distributions by GMM are presented in Tables \ref{micro10efi} and \ref{micro5efi} in the Appendix. Our result suggest that the accuracy does not seem to be strongly affected by the estimation method in the case of the GB2 distribution. However, as observed above, the size of the error increases with the number of overidentifying restrictions. Indeed, GMM estimates of the lognormal distribution seem to present much larger estimation errors than NLS.} Since the two-parameter distributions seem to lead to less reliable estimates of inequality measures, we have only included the results of the lognormal distribution because this model has been conventionally employed to estimate the size distribution of income.\footnote{We have also computed the absolute error in the estimation of different inequality measures for the Fisk and the Weibull distributions. These results are available upon request.} Both the Gini coefficient and the Atkinson index have been estimated by Monte Carlo simulation, from samples of size $N=10^6$.

\begin{table}
\begin{center}

\caption{\label{micro10}
Absolute difference between estimated and observed inequality measures: 10 income shares}
\vspace{0.2cm}
\footnotesize{
\begin{threeparttable}
\begin{tabular}{l l c c c c c c}
\toprule
&		 &	Lower bound	&	GB2	&	Beta 2	&	Singh-Maddala	&	Dagum	&	Log-normal	\\
\midrule
	&	Mean	&	0.0071	&	0.0007	&	0.0018	&	0.0012	&	0.0014	&	0.0022	\\
	&	[0, 0.01)	&	85.97\%	&	99.64\%	&	99.28\%	&	98.92\%	&	98.92\%	&	100\%	\\
Gini	&	[0.01, 0.02)	&	14.03\%	&	0.36\%	&	0.72\%	&	1.08\%	&	1.08\%	&	0\%	\\
Index	&	[0.02, 0.05)	&	0\%	&	0\%	&	0\%	&	0\%	&	0\%	&	0\%	\\
	&	[0.05, 0.1)	&	0\%	&	0\%	&	0\%	&	0\%	&	0\%	&	0\%	\\
	&	[0.1, )	&	0\%	&	0\%	&	0\%	&	0\%	&	0\%	&	0\%	\\
	&	Mean	&	0.0066	&	0.0020	&	0.0040	&	0.0031	&	0.0038	&	0.0045	\\
	&	[0, 0.01)	&	98.56\%	&	98.92\%	&	98.2\%	&	94.96\%	&	88.13\%	&	96.4\%	\\
Atkinson	&	[0.01, 0.02)	&	1.44\%	&	1.08\%	&	1.08\%	&	3.96\%	&	10.79\%	&	3.6\%	\\
index ($\epsilon = 0.5$)	&	[0.02, 0.05)	&	0\%	&	0\%	&	0.72\%	&	1.08\%	&	1.08\%	&	0\%	\\
	&	[0.05, 0.1)	&	0\%	&	0\%	&	0\%	&	0\%	&	0\%	&	0\%	\\
	&	[0.1, )	&	0\%	&	0\%	&	0\%	&	0\%	&	0\%	&	0\%	\\
	&	Mean	&	0.0131	&	0.0049	&	0.0095	&	0.0065	&	0.0072	&	0.0119	\\
	&	[0, 0.01)	&	27.7\%	&	89.93\%	&	59.71\%	&	80.94\%	&	75.18\%	&	47.48\%	\\
Atkinson	&	[0.01, 0.02)	&	67.63\%	&	8.99\%	&	35.61\%	&	17.27\%	&	19.78\%	&	38.13\%	\\
index ($\epsilon = 1$)	&	[0.02, 0.05)	&	4.68\%	&	1.08\%	&	4.68\%	&	1.08\%	&	3.96\%	&	14.03\%	\\
	&	[0.05, 0.1)	&	0\%	&	0\%	&	0\%	&	0.72\%	&	1.08\%	&	0.36\%	\\
	&	[0.1, )	&	0\%	&	0\%	&	0\%	&	0\%	&	0\%	&	0\%	\\
	&	Mean	&	0.0294	&	0.0161	&	0.0246	&	0.0173	&	0.0183	&	0.0297	\\
	&	[0, 0.01)	&	3.96\%	&	44.24\%	&	20.14\%	&	38.49\%	&	42.45\%	&	20.5\%	\\
Atkinson	&	[0.01, 0.02)	&	27.7\%	&	23.74\%	&	28.78\%	&	28.78\%	&	25.54\%	&	20.5\%	\\
index ($\epsilon = 1.5$)	&	[0.02, 0.05)	&	59.71\%	&	30.58\%	&	43.88\%	&	28.42\%	&	25.9\%	&	40.65\%	\\
	&	[0.05, 0.1)	&	8.63\%	&	1.44\%	&	7.19\%	&	3.6\%	&	5.04\%	&	17.27\%	\\
	&	[0.1, )	&	0\%	&	0\%	&	0\%	&	0.72\%	&	1.08\%	&	1.08\%	\\
\bottomrule
\end{tabular}

\begin{tablenotes}
\item Note: Results based on 278 datasets of the LIS database. Parametric models have been estimated by NLS. All inequality measures have been estimated by Monte Carlo simulation using samples of size $N=10^6$.
\end{tablenotes}
\end{threeparttable}
}
\end{center}
\end{table}

\begin{table}[h]
\begin{center}

\caption{\label{micro5}
Absolute difference between estimated and observed inequality measures: 5 income shares}
\vspace{0.2cm}
\footnotesize{
\begin{threeparttable}
\begin{tabular}{l l c c c c c c}
\toprule
	&	&	Lower bound	&	GB2	&	Beta 2	&	Singh-Maddala	&	Dagum	&	Log-normal	\\
\midrule
	&	Mean	&	0.0225	&	0.0013	&	0.0023	&	0.0017	&	0.0019	&	0.0027	\\
	&	[0, 0.01)	&	0\%	&	98.92\%	&	98.92\%	&	98.92\%	&	98.92\%	&	98.92\%	\\
Gini	&	[0.01, 0.02)	&	46.76\%	&	1.08\%	&	0.72\%	&	0.72\%	&	0\%	&	1.08\%	\\
Index	&	[0.02, 0.05)	&	53.24\%	&	0\%	&	0.36\%	&	0.36\%	&	1.08\%	&	0\%	\\
	&	[0.05, 0.1)	&	0\%	&	0\%	&	0\%	&	0\%	&	0\%	&	0\%	\\
	&	[0.1, )	&	0\%	&	0\%	&	0\%	&	0\%	&	0\%	&	0\%	\\
	&	Mean	&	0.0133	&	0.0028	&	0.0043	&	0.0036	&	0.0041	&	0.0048	\\
Atkinson	&	[0, 0.01)	&	26.26\%	&	97.84\%	&	97.84\%	&	92.45\%	&	87.05\%	&	96.76\%	\\
index ($\epsilon = 0.5$)	&	[0.01, 0.02)	&	58.99\%	&	1.8\%	&	1.08\%	&	6.47\%	&	11.87\%	&	2.88\%	\\
	&	[0.02, 0.05)	&	14.75\%	&	0.36\%	&	1.08\%	&	1.08\%	&	1.08\%	&	0.36\%	\\
	&	[0.05, 0.1)	&	0\%	&	0\%	&	0\%	&	0\%	&	0\%	&	0\%	\\
	&	[0.1, )	&	0\%	&	0\%	&	0\%	&	0\%	&	0\%	&	0\%	\\
	&	Mean	&	0.0247	&	0.0055	&	0.0092	&	0.0063	&	0.0064	&	0.0124	\\
	&	[0, 0.01)	&	0\%	&	86.69\%	&	60.07\%	&	82.37\%	&	78.42\%	&	43.53\%	\\
Atkinson	&	[0.01, 0.02)	&	32.37\%	&	12.59\%	&	34.89\%	&	15.83\%	&	17.27\%	&	42.09\%	\\
index ($\epsilon = 1$)	&	[0.02, 0.05)	&	67.63\%	&	0.72\%	&	5.04\%	&	1.8\%	&	4.32\%	&	14.03\%	\\
	&	[0.05, 0.1)	&	0\%	&	0\%	&	0\%	&	0\%	&	0\%	&	0.36\%	\\
	&	[0.1, )	&	0\%	&	0\%	&	0\%	&	0\%	&	0\%	&	0\%	\\
	&	Mean	&	0.0458	&	0.0166	&	0.0234	&	0.0163	&	0.0159	&	0.0302	\\
	&	[0, 0.01)	&	0\%	&	43.53\%	&	20.5\%	&	40.65\%	&	45.68\%	&	17.99\%	\\
Atkinson	&	[0.01, 0.02)	&	4.32\%	&	21.58\%	&	28.42\%	&	27.7\%	&	28.06\%	&	21.22\%	\\
index ($\epsilon = 1.5$)	&	[0.02, 0.05)	&	56.47\%	&	33.09\%	&	45.32\%	&	28.06\%	&	21.22\%	&	42.09\%	\\
	&	[0.05, 0.1)	&	38.49\%	&	1.8\%	&	5.76\%	&	3.6\%	&	5.04\%	&	17.63\%	\\
	&	[0.1, )	&	0.72\%	&	0\%	&	0\%	&	0\%	&	0\%	&	1.08\%	\\
\bottomrule
\end{tabular}

\begin{tablenotes}
\item Note: Results based on 278 datasets of the LIS database. Parametric models have been estimated by NLS. All inequality measures have been estimated by Monte Carlo simulation using samples of size $N=10^6$.
\end{tablenotes}
\end{threeparttable}
}
\end{center}
\end{table}

Our results suggest that all functional forms seem to lead to very accurate estimates of the Gini coefficient, with estimates that differ from the observed value in less than 0.02.  For the Atkinson index, the accuracy of the estimates seems to depend on the value of the inequality aversion parameter. Our results suggest that the estimates of the Atkinson index become less reliable as the value of the parameter increases. The GB2, the Singh-Maddala and the Dagum distributions show accurate estimates of the Atkinson measure for parameter values lower than one. When the sensitivity parameter is larger than 1, the measure is considerably sensitive to the lower end of the distribution, meaning that the value of this inequality measure is largely influenced by the left tail of the distribution. Hence, even if the bulk of the distribution is adequately modeled, relatively small errors in the representation of the left tail might bias the estimates of the Atkinson index. Although the reliability of the estimates is inversely associated with the inequality aversion parameter, the GB2 and the Singh-Madala distributions report relatively accurate estimates of the Atkinson index ($\epsilon = 1.5$), which differ in less than 0.05 in the 98\% of the datasets. 

The comparison of figures in Tables \ref{micro10} and \ref{micro5} reveals that the error in the estimation of inequality measures is slightly larger if the estimates are obtained from 5 data points. However, even with 5 income shares, the GB2 family of income distributions yields reliable estimates of inequality measures, which confirms the results presented in Section \ref{macro} that these models provide us with an excellent methodological framework to estimate income inequality from grouped data, even when only a few points of the Lorenz curve are available.

\subsection{Estimation of income inequality in bimodal distributions}

A great deal of the criticism directed at the use of parametric models to estimate the Lorenz curve from grouped data is the misspecification error that may arise as a consequence of imposing of a particular functional form. Although the GB2 family is acknowledged to be an outstanding candidate to model income variables, it is only able to represent one- and zero-mode distributions. These are the expected shapes of the income distribution in most countries: one mode is conventionally observed in
developed countries with a well-established middle class, while zero mode distributions are
characteristic of developing countries, which present high poverty rates. However, the conjunction of these two factors leads to bimodal distributions, which are typically observed in economies in transition.

Although bimodal distributions are the exception rather than the rule, prior research has repeatedly emphasaised the potential consequences of using parametric models in these cases, thus justifying the use of the lower bound on grounds of reliability and practicability. The parametric approach requires defining \textit{ex-ante} the functional form of the distribution, but grouped data on few points of the Lorenz curve are not informative enough to ascertain the number of modes of the distribution.\footnote{Krause (\citeyear{krause2014}) developed a method to determine whether a distribution is unimodal or zeromodal, but gives no insights on the potential bimodality of the income distribution.} Hence, a conservative strategy is to estimate a general model that fits the regular features of the income distribution, typically unimodal or zeromodal, because it is not possible to determine where to deploy alternative parametric models to better capture the bimodality of income data. In those cases, the GB2 family would approximate the bimodality by unimodal/zeromodal functional forms, thus leading to inaccurate estimates. 

In this section, we examine the size of the error in the estimation of inequality measures using the GB2 family when the income distribution presents a bimodal density function. We use Monte Carlo simulation to obtain synthetic samples from a mixture of a Weibull and a truncated normal distribution with a probability density function (pdf) of the following form:
\begin{equation}\label{mix1}
f(x_1; \beta, \alpha, \omega, \mu, \sigma) = \omega \frac{\beta}{\alpha^{\beta}}x_i^{\beta-1} \exp \left[-\left(\frac{x_i}{\alpha}\right)^{\beta}\right]+ (1-\omega)\frac{\phi(x_i; \mu, \sigma^2)}{\Phi(\mu/\sigma)},
\end{equation}%
where $\omega, 0\leq \omega \leq 1,$ represents the mixing proportion of the Weibull distribution with scale parameter $\alpha$ and shape parameter $\beta$; $\phi(x,  \mu, \sigma^2)$ is the probability density function of a normal distribution with mean $\mu$ and variance $\sigma^2$ and $\Phi()$ represents the cumulative distribution function of the standard normal distribution.

The pdf in (\ref{mix1}) has been used by Paap and van Dijk (\citeyear{paap1998}) to estimate the cross-sectional distribution of income in 120 countries at six periods of time from 1960 to 1989. We rely on their estimates because they are expected to depict bimodal shapes which are particularly representative of income variables. Figure 3 presents the histograms and the kernel density functions of the simulated samples ($N=10000$) for different parameter values. The simulated samples show a variety of shapes of the density function, going from a unimodal distribution with a heavy right tail to a bimodal distribution where the two components of the mixture are clearly identified. 

\begin{figure}\label{distbimodal}
\caption{Histograms and kernel density functions of simulated samples from a mixture of a Weibull and a truncated normal distribution for different parameter values}
\begin{tabular}{c c}
    \includegraphics*[scale=0.35]{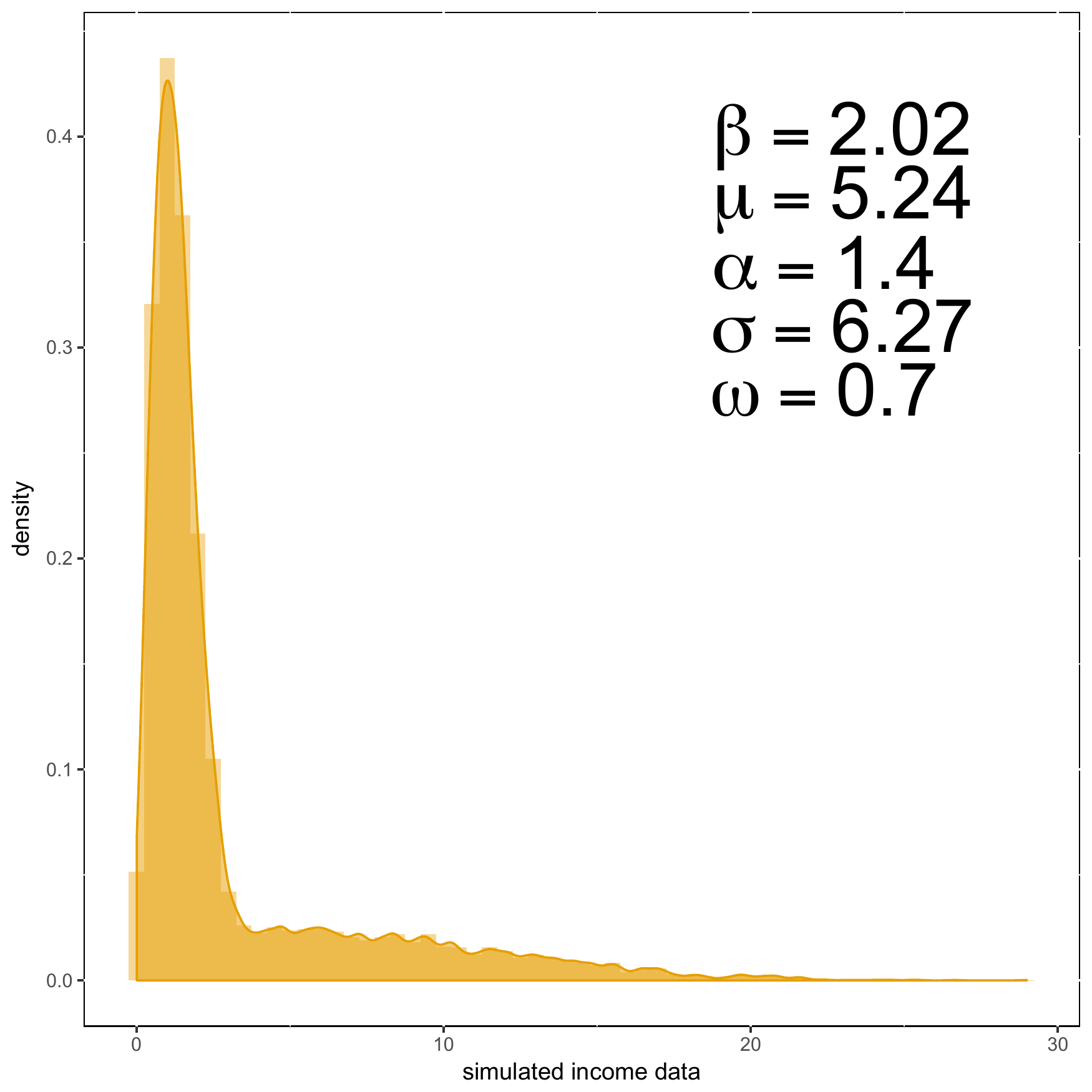}&\includegraphics*[scale=0.35]{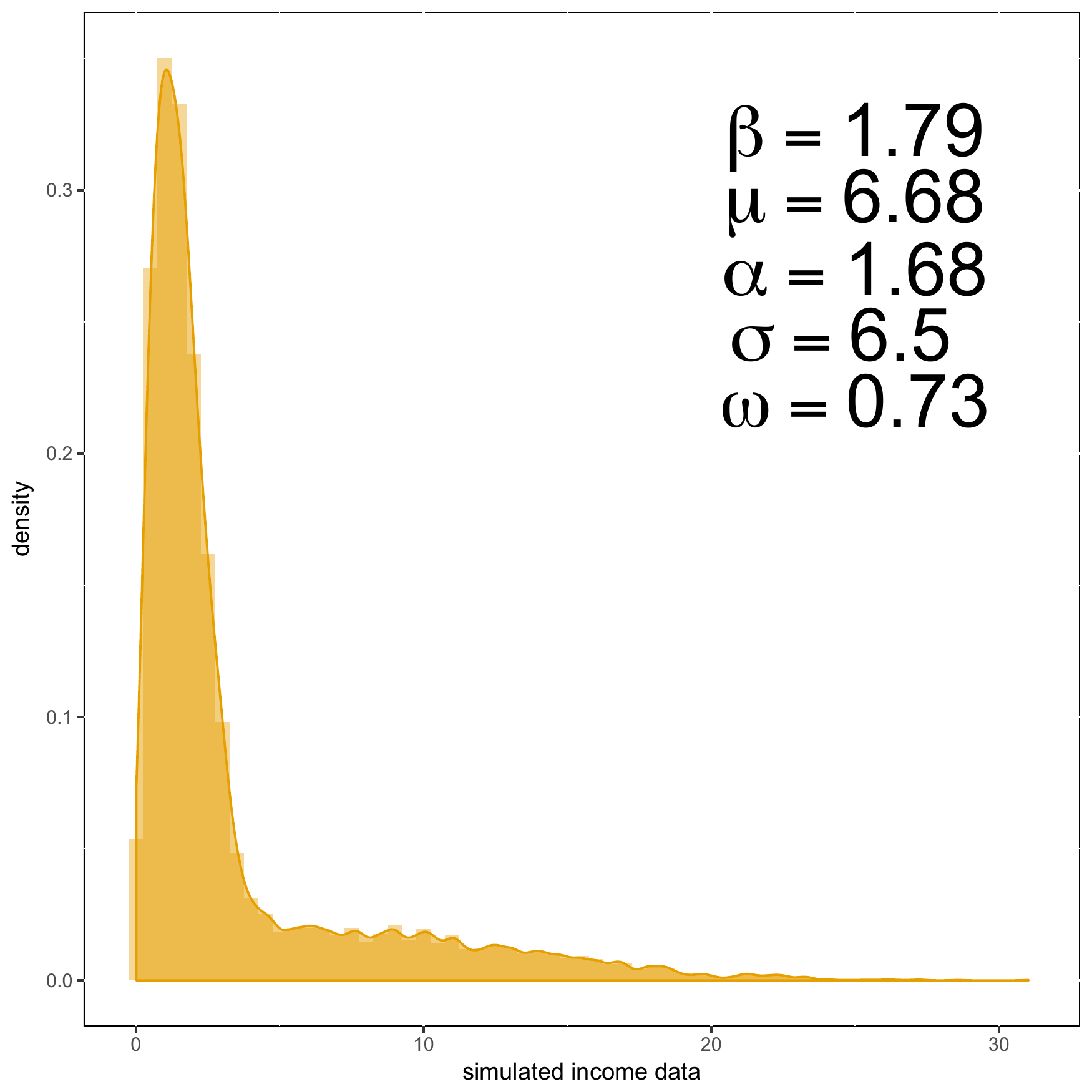}\\
     \includegraphics*[scale=0.35]{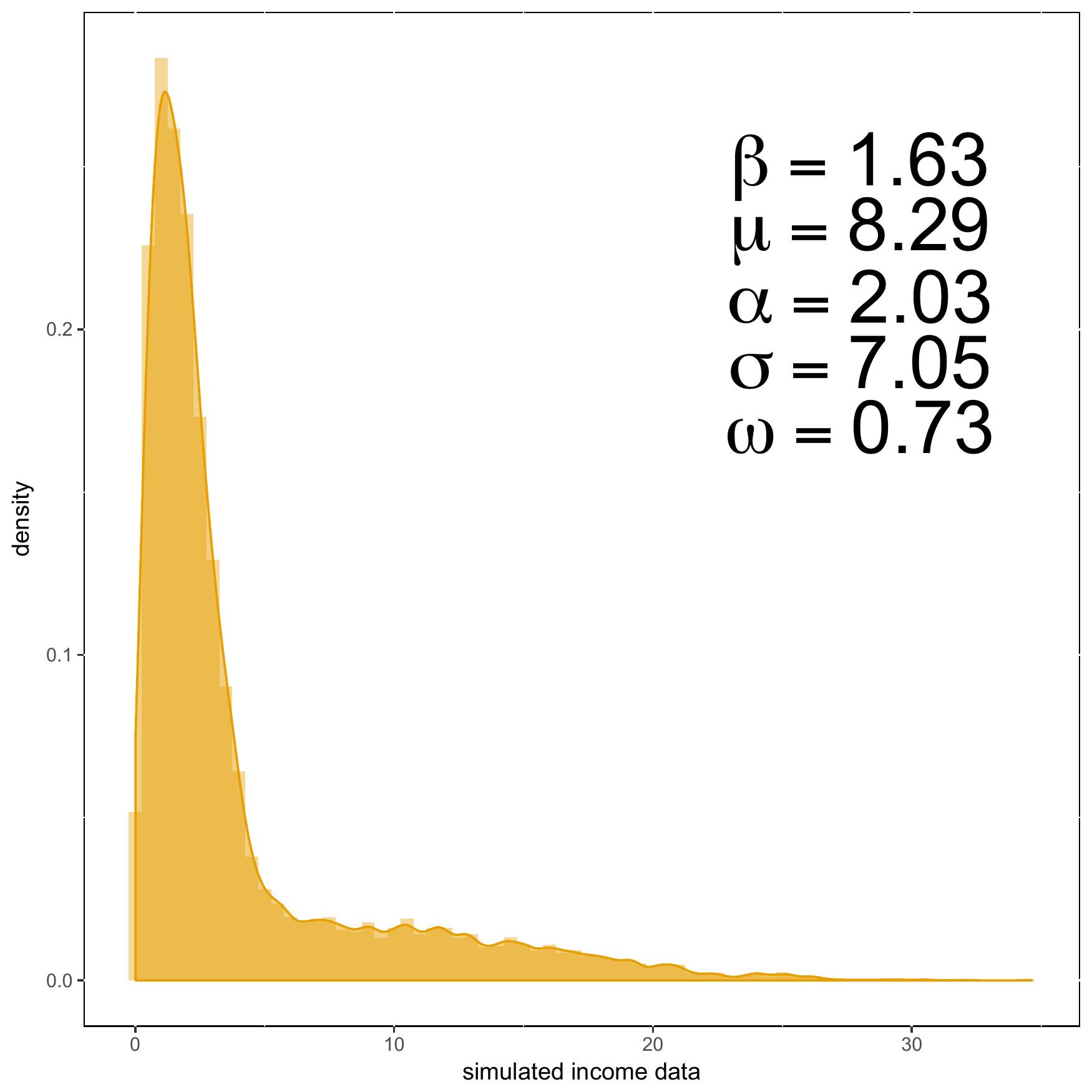}&\includegraphics*[scale=0.35]{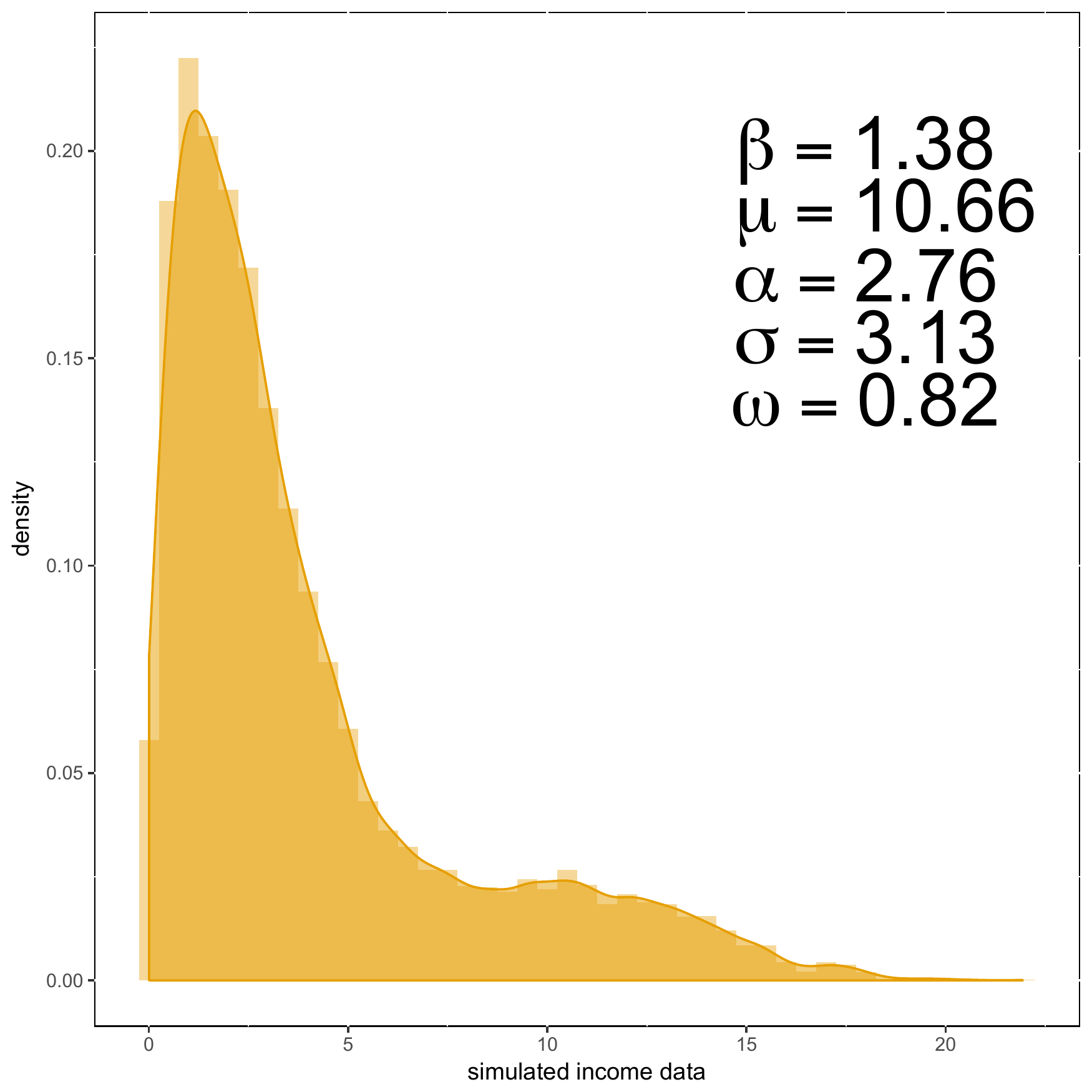}\\
      \includegraphics*[scale=0.35]{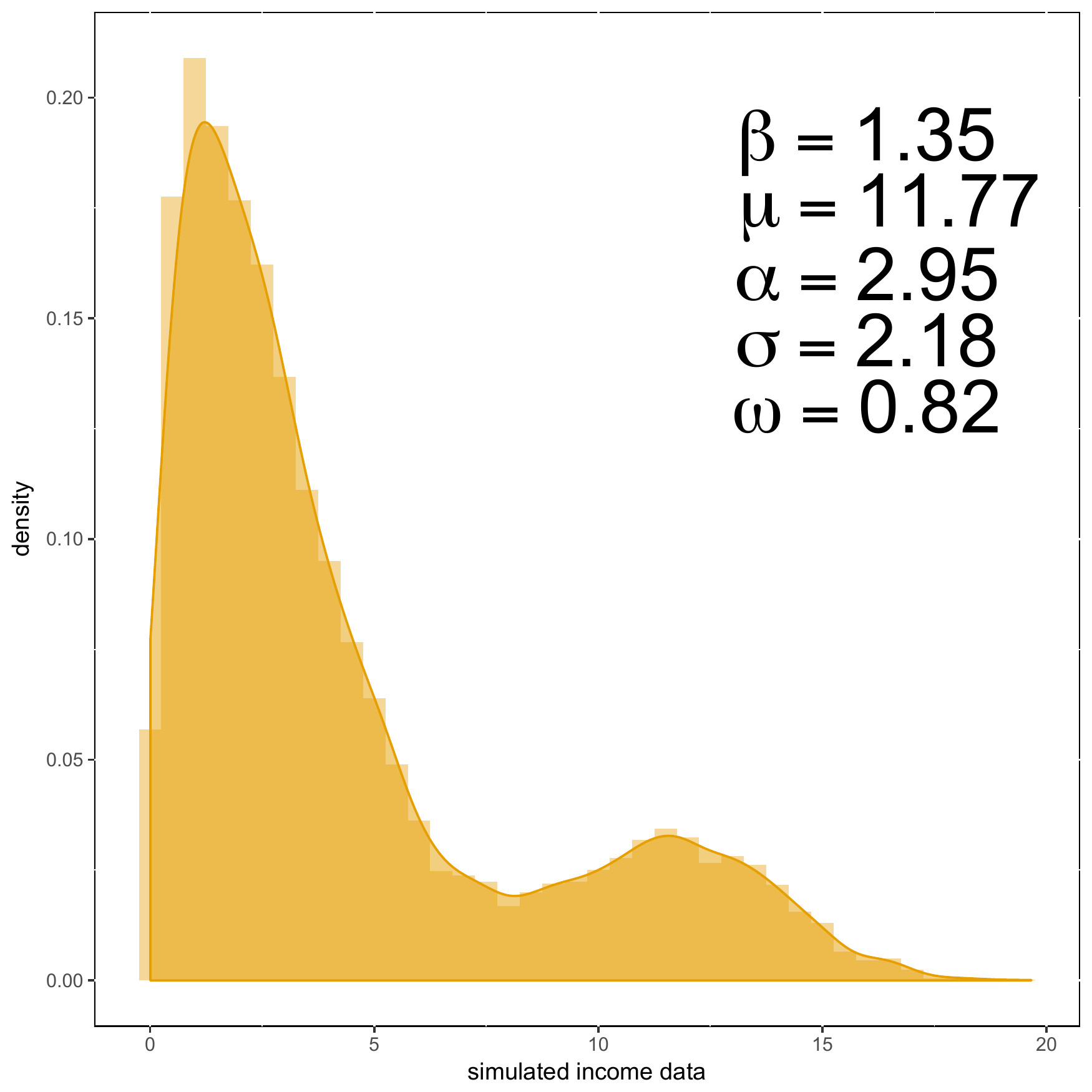}&\includegraphics*[scale=0.35]{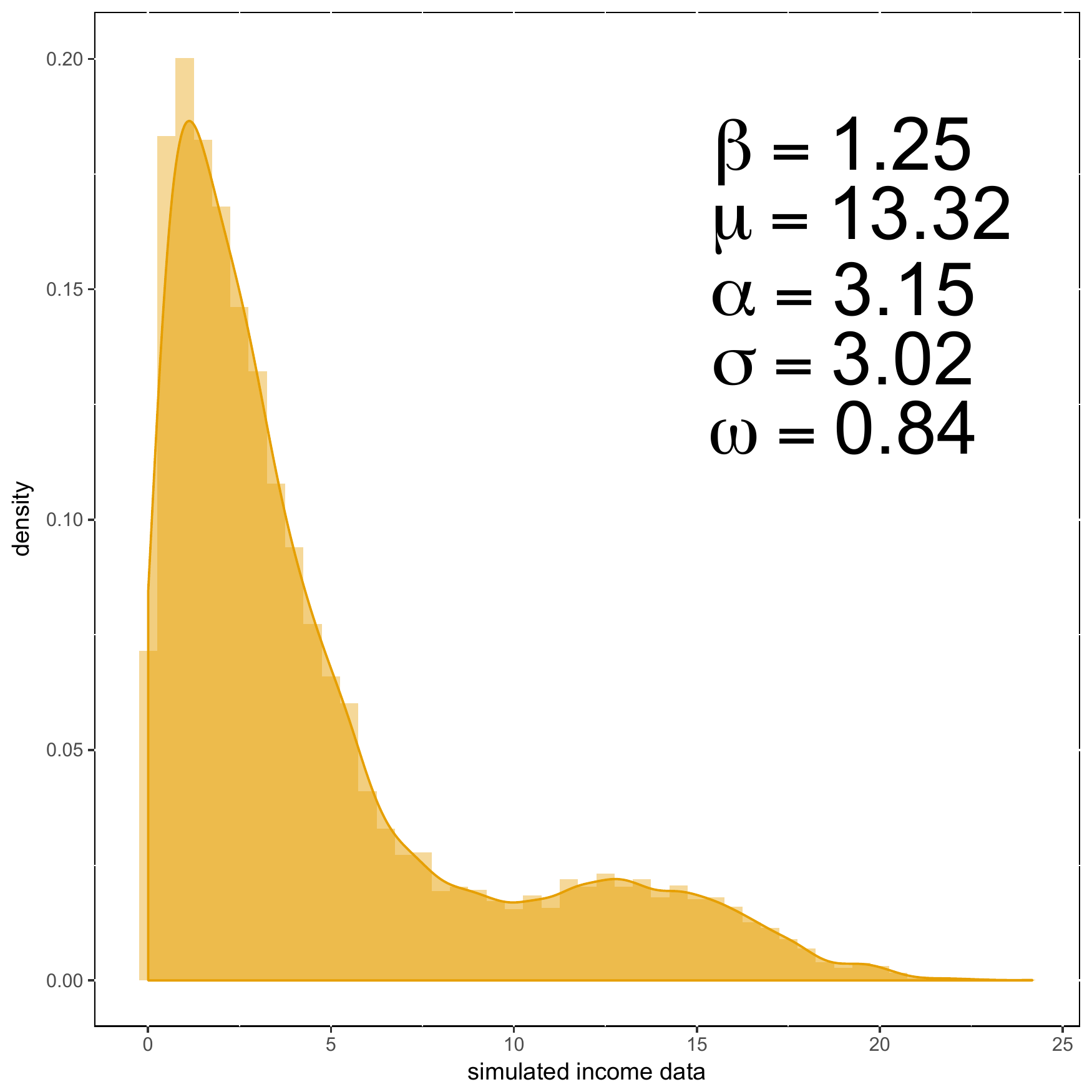}\\
\end{tabular}

\footnotesize{Note: The histograms have been normalised so that the area under the bars is equal to one, making them comparable with the kernel density functions}

\end{figure}

To illustrate the situation in which only grouped data is available, we obtain five and ten income shares from the simulated samples of bimodal distributions to have limited information with the same structure as the WIID.  We also compute the Gini index and the Atkinson measure setting $\epsilon= 0.5, 1, 1.5$. These values of the inequality measures are taken as a benchmark to evaluate the performance of the GB2 family to estimate income inequality in bimodal distributions. The \textit{simulated} grouped data is used to estimate different models of the GB2 family by NLS, deploying the estimation techniques described in Section \ref{estimation}. The corresponding Gini index and Atkinson measures are estimated by Monte Carlo simulation.

To evaluate the size of the error in the estimation of relative inequality measures, we calculate the absolute difference between estimated measures with the ones obtained from the bimodal distributions. A summary of this information is presented in Table \ref{bimodal}, which presents the average error in the estimation of the Gini coefficient and the  Atkinson index for different models of the GB2 family. We have also computed the corresponding lower bound of these measures to analyse whether the nonparametric approach leads to more accurate estimates when the underlying income distribution presents two modes.

Our estimates reveal that the gap between the estimation errors of the parametric and the nonparametric approaches has narrowed substantially with respect to the previous results (Tables \ref{DiffNLS}, \ref{micro10} and \ref{micro5}) which are mostly based on unimodal distributions. As expected, the GB2 seems to provide more accurate estimates of inequality measures than the three-parameter models. Our results also suggest that this model  yields, on average, more accurate estimates of the Gini index than the lower bound. This result is also observed for the Atkinson measures that are very sensitive to the left tail of the distribution. In contrast, when the value of the inequality aversion parameter is low, the lower bound presents more reliable estimates than the GB2 distribution. It should be worth mentioning, however, that any of the estimation techniques leads systematically to more reliable estimates of the inequality measures.\footnote{The complete results for the six simulated samples are available upon request.}

 \begin{table}
\begin{center}
\caption{\label{bimodal}
Average absolute difference between estimated and observed inequality measures: ten and five income shares}
\vspace{0.2cm}
\footnotesize{
\begin{threeparttable}
\begin{tabular}{l c c c c c c}
\toprule
10 income shares	&		&		&		&		&		&		\\
\midrule
	&	Lower bound	&	GB2	&	Beta 2	&	Singh-Maddala	&	Dagum	&	Lognormal	\\
\midrule
Gini index					&	0.0085	&	0.0071	&	0.0082	&	0.0092	&	0.0110	&	0.0057	\\
Atkinson index ($\epsilon=0.5$)	&	0.0053	&	0.0109	&	0.0153	&	0.0180	&	0.0261	&	0.0093	\\
Atkinson index ($\epsilon=1$)	&	0.0132	&	0.0162	&	0.0268	&	0.0349	&	0.0535	&	0.0291	\\
Atkinson index ($\epsilon=1.5$)	&	0.0331	&	0.0260	&	0.0532	&	0.0739	&	0.1332	&	0.0576	\\
\midrule
5 income shares	&		&		&		&		&		&		\\
\midrule
Gini index					&	0.0308	&	0.0181	&	0.0177	&	0.0199	&	0.0213	&	0.0149	\\
Atkinson index ($\epsilon=0.5$)	&	0.0160	&	0.0276	&	0.0258	&	0.0327	&	0.0359	&	0.0148	\\
Atkinson index ($\epsilon=1$)	&	0.0315	&	0.0119	&	0.0117	&	0.0210	&	0.0285	&	0.0293	\\
Atkinson index ($\epsilon=1.5$)	&	0.0610	&	0.0081	&	0.0109	&	0.0148	&	0.0392	&	0.0571	\\
\bottomrule
\end{tabular}

\begin{tablenotes}
\item Note: Results based on simulated samples of size 10000 of mixtures of a Weibull and a normal distribution with the following parameter values: ($\beta$, $\mu$, $\alpha$, $\sigma$, $\omega$)= (2.02, 5.24, 1.4, 6.27, 0.7), (1.79, 6.68, 1.68, 6.5, 0.73), (1.63, 8.29, 2.03, 7.05, 0.73), (1.38, 10.66, 2.76, 3.13, 0.82), (1.35, 11.77, 2.95, 2.18, 0.82), (1.25, 13.32, 3.15, 3.02, 0.84).  All inequality measures have been estimated by Monte Carlo simulation using samples of size $N=10^6$.
\end{tablenotes}
\end{threeparttable}
}
\end{center}
\end{table}

As previously observed, the size of the estimation error using the nonparametric approach increases substantially if the estimation is based on five instead of ten income shares. It is, therefore, surprising that, on average, it still leads to a better approximation of the Atkinson index with $\epsilon = 0.5$ than the GB2 distribution. The estimation of the parametric models is dominated by the first mode, thus providing an accurate fit for the bottom part of the distribution at expense of a relatively poor fit at the right tail of the distribution. Hence, the GB2 tends to yield more accurate estimates to the measures that are particularly sensitive to the left tail of the distribution. 

\section{Conclusions}

Over the past decades, there has been a growing interest in the distributional patterns of income in both, 
the economic literature and the international policy arena. The introduction of the Sustainable Development 
Goals has highlighted the relevance of this topic since  Goal 10 calls for reducing inequalities in income, 
thus positioning disparities as a key concern, not only because wellbeing is a prerogative of all citizens, 
but also because sustained development itself is impeded by high inequalities. Although addressing inequality 
trends has become essential, individual data on income or consumption is not often available. Instead, 
group data from nationally representative surveys are, in most cases, used to assess 
the evolution of inequality levels.

In this context of limited information, most prior research on global inequality relies on lower bounds of inequality measures, constructed under the assumption of equality of incomes within each income group. While being an intuitive method, it obviously leads to biased estimates of inequality measures. To provide reliable results, we must define more plausible assumptions on income dynamics within incomes shares. In this paper, we explore the practical implications of using the GB2 family of distributions to estimate income inequality from grouped data. We first focused on the estimation method by comparing the performance of conventional NLS and optimally-weighted GMM. Our estimates reveal that NLS yields more accurate estimates of the Gini index than GMM in most cases. Therefore, when the priority is to obtain unbiased estimates of inequality measures, NLS should be preferred over GMM, even though it means to sacrifice asymptotic efficiency.

A potential limitation of using parametric models is the requirement to impose a particular functional form to describe the income distribution that could lead to biased estimates in case that the model is not able to represent adequately income dynamics. Indeed, misspecification bias has been the central argument in favor of using the lower bound to estimate inequality from grouped data. To address this issue, we have compared the performance of the GB2 family to estimate different inequality measures to the nonparametric lower bound estimates. Our results suggest that the parametric approach provides much more accurate results than the conventionally used lower bound, which seems to notably underestimate inequality measures in most cases. Even the two-parameter distributions  yield more reliable estimates of inequality measures than the lower bound, although more complex models are generally preferred over the simplest ones. Only for bimodal distributions, the lower bound and the parametric approach report estimates of similar precision.  

Our estimates suggest, therefore, that much of the research on economic inequality relies on severely biased estimates. We show that the GB2 distribution provides an excellent approximation of the income distribution, which yields reliable estimates of relative inequality measures in virtually all cases. We expect that this strong result along with the development of the convenient R package \texttt{GB2group}, which deploys the estimation of this model from grouped data in form of income shares by NLS and GMM, might contribute to incentivize the use of this model and to obtain improved estimates on income inequality.

\section*{Acknowledgements}

Vanesa Jorda and Jose Maria Sarabia acknowledge the financial support of the Ministerio de Econom\'ia y Competitividad (Project ECO2016-76203-C2-1-P).
\newpage

\bibliographystyle{ecta}
\bibliography{ParamModels}

\newpage
\section*{Appendix}

\begin{table}[h]
\begin{center}

\caption{\label{DiffGMM}
Absolute error in the estimation of the Gini index using linear interpolation and different parametric distributions of the GB2 family}
\vspace{0.2cm}
 \small{
\begin{threeparttable}

\begin{tabular}{l c c c c c}
\toprule
Distribution	&	[0, 0.01)	&	[0.01, 0.02)	&	[0.02, 0.05)	&	[0.05, 0.1)	&	[0.1, )	\\
         \midrule
Lower bound	&	70.94\%	&	21.54\%	&	6.12\%	&	1.03\%	&	0.37\%	\\
GB2	&	91.4\%	&	5.5\%	&	2.19\%	&	0.79\%	&	0.12\%	\\
B2	&	81.36\%	&	11.91\%	&	5.5\%	&	1.12\%	&	0.12\%	\\
SM	&	77.39\%	&	10.79\%	&	9.55\%	&	2.07\%	&	0.21\%	\\
Dagum	&	67.22\%	&	17.36\%	&	13.35\%	&	1.94\%	&	0.12\%	\\
Lognormal	&	34.23\%	&	23.07\%	&	34.44\%	&	6.82\%	&	1.45\%	\\
Fisk	&	40.55\%	&	25.8\%	&	28.19\%	&	4.96\%	&	0.5\%	\\
Weibull	&	1.16\%	&	1.57\%	&	24.64\%	&	53.12\%	&	19.51\%	\\ 
\bottomrule
\end{tabular}

\begin{tablenotes}
\item Note: Results based on 3286 datasets of the WIID. Parametric models have been estimated by GMM. For the parametric distributions except the GB2, the Gini coefficient has been estimated using the formulas in Table \ref{GB2gini}. The Gini index of the GB2 distribution has been estimated by Monte Carlo simulation using samples of size $N=10^6$.
\end{tablenotes}
\end{threeparttable}

}
\end{center}
\end{table}

\begin{table}[h]
\begin{center}

\caption{\label{BIC table}
GOF matrix: Bayesian information criterion (BIC)}
\vspace{0.2cm}
 \small{
\begin{threeparttable}

\begin{tabular}{l c c c c c c c}
\toprule
	&	GB2	&	Beta 2	&	Singh-Maddala	&	Dagum	&	Lognormal	&	Fisk	&	Weibull	\\
	\midrule
GB2	&	100\%	&	87\%	&	85\%	&	88\%	&	98\%	&	98\%	&	99\%	\\
Beta 2	&	13\%	&	100\%	&	43\%	&	48\%	&	96\%	&	84\%	&	98\%	\\
Singh-Maddala	&	15\%	&	57\%	&	100\%	&	62\%	&	92\%	&	89\%	&	99\%	\\
Dagum	&	12\%	&	52\%	&	38\%	&	100\%	&	87\%	&	89\%	&	99\%	\\
Lognormal	&	2\%	&	4\%	&	8\%	&	13\%	&	100\%	&	45\%	&	95\%	\\
Fisk	&	2\%	&	16\%	&	11\%	&	11\%	&	55\%	&	100\%	&	91\%	\\
Weibull	&	1\%	&	2\%	&	1\%	&	1\%	&	5\%	&	9\%	&	100\%	\\
\bottomrule
\end{tabular}

\begin{tablenotes}
\item Note: Results based on 5215 datasets of the WIID. Parametric models have been estimated by GMM.
\end{tablenotes} 
\end{threeparttable}
}
\end{center}
\end{table}

\begin{table}[h]
\begin{center}

\caption{\label{Wtable}
GOF matrix: Weighted sum of squared residuals}
\vspace{0.2cm}
 \small{
\begin{threeparttable}

\begin{tabular}{l c c c c c c c}
\toprule
	&	GB2	&	Beta 2	&	Singh-Maddala	&	Dagum	&	Lognormal	&	Fisk	&	Weibull	\\
	\midrule
GB2	&	100\%	&	98\%	&	97\%	&	98\%	&	100\%	&	100\%	&	100\%	\\
Beta 2	&	2\%	&	100\%	&	43\%	&	49\%	&	98\%	&	88\%	&	98\%	\\
Singh-Maddala	&	3\%	&	57\%	&	100\%	&	61\%	&	94\%	&	97\%	&	99\%	\\
Dagum	&	2\%	&	51\%	&	39\%	&	100\%	&	89\%	&	98\%	&	99\%	\\
Lognormal	&	0\%	&	2\%	&	6\%	&	11\%	&	100\%	&	45\%	&	95\%	\\
Fisk	&	0\%	&	12\%	&	3\%	&	2\%	&	55\%	&	100\%	&	91\%	\\
Weibull	&	0\%	&	2\%	&	1\%	&	1\%	&	5\%	&	9\%	&	100\%	\\
\bottomrule
\end{tabular}

\begin{tablenotes}
\item Note: Results based on 5215 datasets of the WIID. Parametric models have been estimated by GMM.
\end{tablenotes}
\end{threeparttable}
}
\end{center}
\end{table}

\begin{table}[h]
\begin{center}

\caption{\label{micro10efi}
Absolute difference between estimated and observed inequality measures: 10 income shares}
\vspace{0.2cm}
\footnotesize{
\begin{threeparttable}
\begin{tabular}{l l c c c c c}
\toprule
Inequality measure	&	Absolute difference	&	GB2	&	Beta 2	&	Singh-Maddala	&	Dagum	&	Lognormal	\\
\midrule
	&	[0, 0.01)	&	99.64\%	&	96.04\%	&	92.09\%	&	79.5\%	&	43.88\%	\\
Gini	&	[0.01, 0.02)	&	0.36\%	&	3.6\%	&	37.41\%	&	8.27\%	&	27.34\%	\\
Index	&	[0.02, 0.05)	&	0\%	&	0.36\%	&	34.53\%	&	11.15\%	&	36.33\%	\\
	&	[0.05, 0.1)	&	0\%	&	0\%	&	32.01\%	&	1.08\%	&	9.35\%	\\
	&	[0.1, )	&	0\%	&	0\%	&	32.01\%	&	0\%	&	6.12\%	\\
Atkinson	&	[0, 0.01)	&	97.84\%	&	96.4\%	&	88.49\%	&	80.94\%	&	71.22\%	\\
index ($\epsilon = 0.5$)	&	[0.01, 0.02)	&	2.16\%	&	2.16\%	&	38.13\%	&	5.4\%	&	27.34\%	\\
	&	[0.02, 0.05)	&	0\%	&	1.44\%	&	37.41\%	&	12.23\%	&	11.51\%	\\
	&	[0.05, 0.1)	&	0\%	&	0\%	&	32.01\%	&	0.36\%	&	7.19\%	\\
	&	[0.1, )	&	0\%	&	0\%	&	32.01\%	&	1.08\%	&	5.76\%	\\
	&	[0, 0.01)	&	90.65\%	&	69.42\%	&	82.37\%	&	73.02\%	&	62.59\%	\\
Atkinson	&	[0.01, 0.02)	&	9.35\%	&	28.78\%	&	46.04\%	&	12.95\%	&	28.06\%	\\
index ($\epsilon = 1$)	&	[0.02, 0.05)	&	0\%	&	1.8\%	&	35.61\%	&	12.59\%	&	19.42\%	\\
	&	[0.05, 0.1)	&	0\%	&	0\%	&	32.01\%	&	0.72\%	&	7.19\%	\\
	&	[0.1, )	&	0\%	&	0\%	&	32.01\%	&	0.72\%	&	5.76\%	\\
	&	[0, 0.01)	&	47.12\%	&	27.34\%	&	58.63\%	&	43.17\%	&	53.96\%	\\
Atkinson	&	[0.01, 0.02)	&	25.18\%	&	30.22\%	&	50.72\%	&	25.9\%	&	33.81\%	\\
index ($\epsilon = 1.5$)	&	[0.02, 0.05)	&	26.26\%	&	39.57\%	&	53.24\%	&	26.98\%	&	23.38\%	\\
	&	[0.05, 0.1)	&	1.44\%	&	2.88\%	&	33.45\%	&	3.96\%	&	6.12\%	\\
	&	[0.1, )	&	0\%	&	0\%	&	32.01\%	&	0\%	&	5.76\%	\\
\bottomrule
\end{tabular}

\begin{tablenotes}
\item Note: Results based on 278 datasets of the LIS database. Parametric models have been estimated by GMM. All inequality measures have been estimated by Monte Carlo simulation using samples of size $N=10^6$.
\end{tablenotes}
\end{threeparttable}
}
\end{center}
\end{table}

\begin{table}[h]
\begin{center}

\caption{\label{micro5efi}
Absolute difference between estimated and observed inequality measures: 5 income shares}
\vspace{0.2cm}
\footnotesize{
\begin{threeparttable}
\begin{tabular}{l l c c c c c}
\toprule
Inequality measure	&	Absolute difference	&	GB2	&	Beta 2	&	Singh-Maddala	&	Dagum	&	Lognormal	\\
\midrule
	&	[0, 0.01)	&	98.56\%	&	98.92\%	&	92.45\%	&	83.09\%	&	58.27\%	\\
Gini	&	[0.01, 0.02)	&	1.44\%	&	0.72\%	&	4.68\%	&	10.07\%	&	24.46\%	\\
Index	&	[0.02, 0.05)	&	0\%	&	0.36\%	&	3.24\%	&	8.99\%	&	15.83\%	\\
	&	[0.05, 0.1)	&	0\%	&	0\%	&	0.72\%	&	1.08\%	&	1.44\%	\\
	&	[0.1, )	&	0\%	&	0\%	&	0.36\%	&	1.08\%	&	0\%	\\
Atkinson	&	[0, 0.01)	&	96.4\%	&	97.84\%	&	88.49\%	&	82.01\%	&	85.97\%	\\
index ($\epsilon = 0.5$)	&	[0.01, 0.02)	&	2.88\%	&	1.08\%	&	7.55\%	&	7.91\%	&	11.87\%	\\
	&	[0.02, 0.05)	&	0.72\%	&	1.08\%	&	3.96\%	&	11.87\%	&	0.72\%	\\
	&	[0.05, 0.1)	&	0\%	&	0\%	&	1.08\%	&	1.44\%	&	1.44\%	\\
	&	[0.1, )	&	0\%	&	0\%	&	0.36\%	&	1.08\%	&	0\%	\\
	&	[0, 0.01)	&	84.53\%	&	60.07\%	&	80.22\%	&	74.46\%	&	71.94\%	\\
Atkinson	&	[0.01, 0.02)	&	15.11\%	&	34.89\%	&	14.39\%	&	14.39\%	&	24.1\%	\\
index ($\epsilon = 1$)	&	[0.02, 0.05)	&	0.36\%	&	5.04\%	&	5.4\%	&	13.31\%	&	2.52\%	\\
	&	[0.05, 0.1)	&	0\%	&	0\%	&	1.08\%	&	1.08\%	&	1.44\%	\\
	&	[0.1, )	&	0\%	&	0\%	&	0.36\%	&	1.08\%	&	0\%	\\
	&	[0, 0.01)	&	42.81\%	&	20.5\%	&	40.65\%	&	41.37\%	&	28.78\%	\\
Atkinson	&	[0.01, 0.02)	&	23.02\%	&	28.42\%	&	26.62\%	&	31.29\%	&	29.5\%	\\
index ($\epsilon = 1.5$)	&	[0.02, 0.05)	&	32.37\%	&	45.32\%	&	29.86\%	&	25.54\%	&	40.29\%	\\
	&	[0.05, 0.1)	&	1.8\%	&	5.76\%	&	3.96\%	&	5.04\%	&	1.44\%	\\
	&	[0.1, )	&	0\%	&	0\%	&	0.36\%	&	1.08\%	&	0\%	\\

\bottomrule
\end{tabular}

\begin{tablenotes}
\item Note: Results based on 278 datasets of the LIS database. Parametric models have been estimated by GMM. All inequality measures have been estimated by Monte Carlo simulation using samples of size $N=10^6$.
\end{tablenotes}
\end{threeparttable}
}
\end{center}
\end{table}

\end{document}